\documentclass[11pt]{article}
\pdfoutput=1

\usepackage[utf8]{inputenc}
\usepackage{multirow}
\usepackage{amsmath, amsfonts, amssymb}
\usepackage{comment}
\usepackage{graphicx}
\usepackage{psfrag}
\usepackage{amsthm}
\usepackage[usenames,dvipsnames,svgnames,table]{xcolor}
\usepackage{enumerate}
\usepackage{arydshln}
\usepackage{soul}
 \usepackage{slashed}
\usepackage{subfig}
\usepackage{mathrsfs}
 \usepackage{a4wide}
  \usepackage{tikz}
  \usepackage{tikz-cd}
  \usetikzlibrary{shapes.geometric}
  \usepackage{tcolorbox}

  \usepackage{color}
  \definecolor{dark-gray}{gray}{0.20}
  \definecolor{gray}{gray}{0.30}
  \definecolor{light-gray}{gray}{0.80}
  \definecolor{dark-red}{rgb}{0.7,0,0}
  \definecolor{dark-green}{rgb}{0.1,0.4,0}
  \definecolor{dark-blue}{rgb}{0.3,0.3,0.7}
  \definecolor{light-blue}{rgb}{0.8,0.8,1}
      \definecolor{swamp}{RGB}{240, 199, 197}
       \definecolor{landscape}{RGB}{180, 250, 199}
          \definecolor{undecided}{RGB}{252, 252, 197}

  \usepackage{pifont}

\usepackage{newunicodechar} 
\newunicodechar{ỳ}{\`y}

\usepackage{setspace}

\newcommand{\beq}{\begin{equation}}  \newcommand{\eeq}{\end{equation}}
\newcommand{\bal}{\begin{aligned}}   \newcommand{\eal}{\end{aligned}}
\newcommand{\be}{\begin{equation}}
\newcommand{\ee}{\end{equation}}
\newcommand{\eq}[1]{(\ref{#1})}

\def\be{\begin{equation}}
\def\ee{\end{equation}}
\def\bea{\begin{eqnarray}}
\def\eea{\end{eqnarray}}

\def\simleq{\; \raise0.3ex\hbox{$<$\kern-0.75em
      \raise-1.1ex\hbox{$\sim$}}\; }
   \def\simgeq{\; \raise0.3ex\hbox{$>$\kern-0.75em
      \raise-1.1ex\hbox{$\sim$}}\; }

\numberwithin{equation}{section}

\usepackage{jheppub}
\usepackage{hyperref}
\usepackage{cleveref}

\hypersetup{
	colorlinks=true,
	linkcolor=dark-blue,
	citecolor=dark-red,
	urlcolor=dark-green,
	linktoc=page
}

\theoremstyle{remark}

\crefname{appendix}{Appendix}{Appendices}

\title{\centering Quantum corrections to DGKT\\ \hspace{0.55cm} and the Weak Gravity Conjecture {\color{white}!?}}

\author{Miguel Montero$^1$,}\affiliation{$^1$Instituto de F\'{i}sica Te\'{o}rica IFT-UAM/CSIC,
C/ Nicol\'{a}s Cabrera 13-15, Campus de Cantoblanco, 28049 Madrid, Spain}
\author{Irene Valenzuela$^{1,2}$}\affiliation{$^2$CERN, Theoretical Physics Department, 1211 Meyrin, Switzerland}

\emailAdd{miguel.montero@csic.es}
\emailAdd{irene.valenzuela@cern.ch}

\abstract{We study D4 brane domain walls in the scale-separated 4d $\mathcal{N}$=1 AdS$_4$ DGKT scenario. Classically, these are BPS and satisfy a no-force condition since their tension equals their charge. We show that this property is not stable against quantum corrections and that these increase the brane tension-to-charge ratio, rendering the branes self-attractive. As a result, DGKT seems to be in tension with the Weak Gravity Conjecture for membranes. The quantum effects we consider include non-perturbative gaugino condensation on the D4-brane worldvolume and Euclidean D2 brane instantons, which correct the tension-to-charge ratio because the DGKT construction breaks all parity symmetries. Similar results hold in other 4d $\mathcal{N}$=1 setups not protected by parity symmetries.}

\begin{document}
\hypersetup{pageanchor=false}
\makeatletter
\let\old@fpheader\@fpheader
\preprint{\begin{flushright} IFT-24-155   \\  CERN-TH-2024-190\end{flushright}}

\makeatother

\maketitle

\hypersetup{
    pdftitle={DGKT},
    pdfauthor={Miguel Montero, Irene Valenzuela},
    pdfsubject={Swampland and Scale Separation}
}

\newcommand{\remove}[1]{\textcolor{red}{\sout{#1}}}

\section{Introduction}

Minimally supersymmetric vacua comprise a fascinating and extremely rich corner of the String Landscape. They enjoy all the virtues of supersymmetry while also allowing for realistic interactions and matter content. For this reason, they often act as basecamps for our attempts to realize our own Universe as a String Theory vacuum. 

On the other hand, these vacua are much harder to control than their higher supersymmetric cousins -- due to the small amount of supersymmetry, there are few quantities protected against quantum corrections.

This paper focuses on quantum corrections in a particularly interesting  minimally supersymmetric vacuum -- the DGKT solution of \cite{DeWolfe:2005uu} (see \cite{Coudarchet:2023mfs} for a recent review; this solution was also simultaneously found in \cite{Camara:2005dc}), and the interplay of these corrections with Swampland constraints (see \cite{Vafa:2005ui,Palti:2020qlc,vanBeest:2021lhn} for reviews). The DGKT vacuum is a four-dimensional anti de Sitter (AdS) solution of the 4d $\mathcal{N}=1$ effective field theory which arises from dimensional reduction of massive IIA string theory on a Calabi-Yau manifold with fluxes. Among its many remarkable properties, the solution seems to exhibit scale separation -- the Kaluza-Klein scale of the compactification manifold is much smaller than the AdS curvature scale, so that the DGKT solution is effectively four-dimensional. This property is not shared by any of the fully controlled top-down AdS vacua known so far, due to their extended supersymmetry, and makes the DGKT much more similar to our own  four-dimensional Universe. Partially for this reason, scale-separated AdS supersymmetric solutions (of which DGKT is the simplest example) are a crucial ingredient in popular strategies to realize de Sitter vacua in string theory such as \cite{Kachru:2003aw,Balasubramanian:2005zx}.

In spite of their importance and mature age, there are a significant number of shadows cast upon the DGKT solution; see \cite{Coudarchet:2023mfs} for an excellent summary. Most of these stem from the fact that DGKT is a solution of a four-dimensional effective field theory, and a fully backreacted ten-dimensional solution has not been constructed yet. Since there are fluxes warping the geometry, the metric is not Calabi-Yau, and hence, no theorem guaranteeing that the ten-dimensional equations of motion have a solution. To this one must add the complications caused by the existence of O6 planes where the metric is not smooth and other subtleties. The solution is also funny when analyzed from the point of view of holography \cite{Banks:2006hg,Aharony:2008wz}; it leads to a holographic dual with a quickly growing central charge, much faster than any controlled top-down examples, a large classical moduli space, and to exactly integer dimensions in the infinite $N$ limit \cite{Conlon:2021cjk} whose CFT interpretation is yet to be found. Understanding all these issues constitutes a very active area of research.

In this paper, we will confront the DGKT scenario with the membrane version of the Weak Gravity Conjecture (WGC) \cite{ArkaniHamed:2006dz}, one of the best established Swampland constraints. The membrane WGC demands the existence of codimension-1 branes whose tension-to-charge ratio is equal to or lower than one in Planck units (see \cite{Aharony:2008wz} for an earlier, classical analysis), a property used in \cite{Ooguri:2016pdq} to argue that non-supersymmetric AdS solutions are always unstable. For a supersymmetric AdS solution, the only way to satisfy the membrane WGC is via exactly extremal -- or BPS -- branes, whose tension equals their charge.\footnote{In the presence of massless scalar fields, the condition for extremality and for balance of forces (saturated by BPS objects) are not exactly the same. In those cases, one has to distinguish between the WGC and the Repulsive Force Condition (RFC) \cite{Palti:2017elp,Heidenreich:2019zkl,Lanza:2020qmt}. However, for codimension one objects, there is no notion of extremality, so the WGC is commonly identified with the condition of having either a BPS or a self-repulsive brane.}

However, in a 4d $\mathcal{N}=1$ vacuum, a BPS membrane is 3d $\mathcal{N}=1$, and this is generically too little supersymmetry to protect the brane tension-to-charge ratio against quantum corrections. The only known way to protect this quantity is when the vacuum preserves a parity symmetry \cite{Bashmakov:2018wts,Gaiotto:2018yjh,Choi:2018ohn}. Otherwise, quantum corrections are expected to generate a non-vanishing potential on the worldvolume field theory of the membrane, which implies a positive contribution to the tension-to-charge ratio; rendering the membrane self-attractive and non-BPS. As we will see, no parity symmetry is preserved in the DGKT vacuum, and we show by explicit computation that these expected corrections do indeed exist, with the result that no exactly BPS branes survive. The DGKT scenario is then in strong tension with the membrane WGC -- both cannot be true at the same time. Irrespectively of which one we have to give up, we are bound to find interesting consequences for holography, as we discuss in the Conclusions section. Moroever, even if we focus on the DGKT scenario, the argument is general and similar problems are expected to arise in any 4d $\mathcal{N}=1$ AdS vacua without parity symmetries. This suggests that the WGC for codimension one branes may only be compatible with AdS vacua whose CFT dual exhibits an exact moduli space protected by symmetries.

The paper is organized as follows: Section \ref{sec:s2}  contains a brief review of the DGKT construction and some of the issues discussed in the literature. Section \ref{sec:s3} introduces the moduli space of the dual field theory due to brane probes of the geometry, and discusses the role of parity symmetries to protect it from quantum corrections , as well as the interplay with the Weak Gravity Conjecture. Section \ref{sec:modquantum} is devoted to the study of this brane moduli space in the DGKT scenario and the explicit computation of the corrections to the worldvolume potential for any location of the brane. Finally, Section \ref{sec:s5} contains some final remarks.

\section{Review of the DGKT construction}\label{sec:s2}
We will start with a short review of the DGKT construction \cite{DeWolfe:2005uu,Camara:2005dc} for a scale separated vacuum. The reader is encouraged to peruse the excellent review \cite{Coudarchet:2023mfs} for more details.

\subsection{The DGKT construction}\label{sec:revw}
The vacuum commonly known in the literature as ``DGKT''  was first constructed as a solution of the 4d $\mathcal{N}=1$ EFT describing the low-energy limit of type II flux compactifications to four dimensions \cite{DeWolfe:2005uu}. The acronym ``DGKT'' refers to the initials of the authors of \cite{DeWolfe:2005uu} and has stuck in the literature, but it is worth noting that these vacua were also analyzed in \cite{Camara:2005dc}, which appeared almost simultaneously. 

The DGKT solution is a flux compactification of massive IIA string theory. In ten-dimensional language, the construction involves a Calabi-Yau manifold with small $h_{2,1}$ (with the rigid case $h_{2,1}=0$, a rigid Calabi-Yau, being ideal) which is then orientifolded \cite{Dabholkar:1997zd};  more specifically, the type IIA ``parent'' CY$_3$ is quotiented by an orientation-reversing involution $\iota_3$, combined with an action of $\Omega (-1)^{F_L}$, which is required for consistency of the worldsheet description. The resulting construction is said to have $O6$ planes at the fixed loci. At energies much below the compactification scale, one expects the dynamics to be captured by the 4d $\mathcal{N}=1$ effective field theory that comes out of the dimensional reduction of the ten-dimensional supergravity to four dimensions, and which was constructed in \cite{Grimm:2004ua}. The fields expected to be relevant are obtained by first performing the standard reduction on the $CY_3$ without fluxes (and which furnish $\mathcal{N}=2$ multiplets), and then keeping only those which are invariant under the orientifold involution. Doing this, one obtains: \begin{itemize}
 \item At the $\mathcal{N}=2$ level, there are  $h_{1,1}$  vector multiplets, where the vector comes from reduction of $C_3$ on the $h_{1,1}$ 2-cycles and the complex scalar comes from the Kahler moduli of the Calabi-Yau (real part) and the reduction of $B_2$ on these two-cycles (imaginary part). The 2-cycles are then divided into $h_{1,1}^+$ orientifold-even cycles and $h_{1,1}^-$ odd cycles. Since $B_2$ and $C_3$ are respectively odd and even under the orientifold action, we get $h_{1,1}^+$  $\mathcal{N}=1$ vector multiplets (containing one vector) and $h_{1,1}^-$ chiral multiplets (with one complex scalar) from this sector. The $h_{1,1}^+$ orientifold-even Kahler modes parametrize Kahler deformations that are incompatible with the orientifold involution, since $\iota_3^*J=-J$, and so, a deformation of the Kahler class $J\,\rightarrow\, J+\delta J$ only commutes with $\iota_3$ if 
 \begin{equation} \iota_3^*(J+\delta J)=-J-\delta J\quad\Rightarrow\quad \iota_3^*\delta J=-\delta J.\end{equation}
 Therefore, these modes are projected out, as required by supersymmetry.
 \item At the $\mathcal{N}=2$ level, there are also $h_{2,1}+1$ $\mathcal{N}=2$ hypermultiplets, each carrying four complex scalars. In $h_{2,1}$ of these, the scalars come from the $h_{2,1}$ complex structure deformations of the CY$_3$ (each of which is itself a complex number), together with an axionic scalar coming from the periods of $C_3$ on the corresponding $h_{2,1}$ cohomology classes (of which there are also two, corresponding to the real and imaginary part of the cohomology class).  Additionally, there is the so-called ``universal'' hypermultiplet \cite{Strominger:1997eb}, whose scalars are given by the volume of the Calabi-Yau, the period of $B_6$ on it, and the periods of $C_3$ along the holomorphic 3-form living in $h_{3,0}$. After taking into account the orientifold, each of these hypermultiplets sees two of its scalars projected out, producing $h_{2,1}+1$ chiral multiplets. The real part of these corresponds to integrals of the complex structure that are compatible with the orientifold involution, and the imaginary part comes from periods of $C_3$ on orientifold-even cycles. 
 \end{itemize}
The $O6$ plane also are magnetic sources of $F_2$ charge, and so they cause an $F_2$ tadpole to appear in the compactification \cite{DeWolfe:2005uu}. This tadpole is then cancelled by turning on $H_3$ flux and Roman's mass, which also contribute to the tadpole due to the ten-dimensional coupling 
 \begin{equation} \mathcal{L}_{10d}\supset -\frac14 m \int C_7\wedge H_3.\end{equation}
Including the contribution from these fluxes, the $F_2$ Bianchi identity becomes
\begin{equation} dF_2=\delta_{O6}+m\, H_3.\label{tad2}\end{equation}
The $H_3$ flux lives in in a 3-cycle which is odd under the orientifold involution; since the field $H_3$ itself is odd under the orientifold \cite{Dabholkar:1997zd}, the flux is invariant and survives the projection. 

Finally, a third flux, 
\begin{equation} N=\int F_4\end{equation}
is turned on along orientifold-even four-cycles. This is compatible with the orientifold because $C_3$ is even under $\Omega (-1)^{F_L}$. This flux does not appear in any tadpole, and in this sense, it is similar to the $F_5$ flux in familiar holographic solutions such as IIB $AdS_5\times S^5$.  Just like in that case, we can take any $N$ we like, and in particular, we can consider the large $N$ limit.

 Once all of this is done, one then computes and solves the equations of motion of the corresponding low-energy 4d $\mathcal{N}=1$ effective field theory with the aforementioned fluxes turned on. This is not an easy task and we refer to \cite{DeWolfe:2005uu} for details. What matters is that, after the dust settles, it turns out that the 4d $\mathcal{N}=1$ EFT admits a supersymmetric solution, where the AdS length scale and a naive estimate of the KK length scale grow as
\begin{equation}\frac{\ell_{\text{AdS}}}{\ell_{\text{P}}}\sim\, N^{\frac94},\quad \frac{\ell_{\text{KK}}}{\ell_{\text{P}}}\sim N^{\frac74},\quad \frac{\ell_{\text{KK}}}{\ell_{\text{AdS}}}\sim\, N^{-\frac12}. \label{estimates}\end{equation}

The ten-dimensional dilaton in turn goes as $ N^{-3/4}$, ensuring that the solution is weakly coupled, and the string length scale goes like $N^{3/2}$ in Planck units, which naively seems to suggest that higher derivative corrections are arbitrarily small.  The moduli described above all get positive masses, save for the $h_{2,1}$ $C_3$-axions, which are massless. In principle, since these directions are exactly massless, sub-leading corrections to the scalar potential in \cite{DeWolfe:2005uu} become important, and are expected to generate a potential with a minimum in some particular value of the axions. Since the axion field is assumed to be compact, there is no danger of a runaway behavior, and the stabilization of these scalars is guaranteed. Nevertheless,  these axions are absent in cases with $h_{2,1}=0$ (rigid Calabi-Yau's), which is why they are preferred in the DGKT literature. And since $\ell_{\text{AdS}}/\ell_{\text{KK}}\rightarrow0$ as $N\rightarrow\infty$, the size of the extra dimensions can be tuned to arbitrarily small values compared to the AdS length scale.  It seems that we have therefore obtained a supersymmetric, scale-separated AdS vacuum. 

We have phrased the above discussion in a way that emphasizes the fact that the DGKT construction is not specific to any given Calabi-Yau, and in principle, it works the same in every case provided that a suitable orientifold involution can be found. However, \cite{DeWolfe:2005uu} also gave a particular concrete realization of the construction in a specific rigid orbifold Calabi-Yau, which we will take as a ``benchmark'' for the whole DGKT scenario.  As described quite detailedly in \cite{DeWolfe:2005uu}, which we follow, the Calabi-Yau orbifold is constructed as a quotient 
\begin{equation}\frac{T^6}{\mathbb{Z}_3\times \mathbb{Z}_3}.\label{cy3}\end{equation}
The coordinates on $T^6$ are taken to be a triplet of complex numbers $(z_1,z_2,z_3)$ subject to identifications
\begin{equation}z_i\sim z_i+1\sim z_i+e^{2\pi i/6}.\end{equation}
The two $\mathbb{Z}_3$ isometries we quotient by in \eq{cy3} are generated by the transformations
\begin{align} (z_1,z_2,z_3)&\rightarrow\, \left( \alpha^2 z_1+\frac{1+\alpha}{3}, \alpha^4 z_1+\frac{1+\alpha}{3}, z_3+\frac{1+\alpha}{3}\right),\nonumber\\(z_1,z_2,z_3)&\rightarrow\,  \alpha^2 (z_1,z_2,z_3),\quad \alpha\equiv e^{2\pi i/6}.\label{trolo}\end{align}
The resulting orbifold Calabi-Yau threefold has nine fixed points, each of which is locally a singularity of the form $\mathbb{C}^3/\mathbb{Z}_3$ where the $\mathbb{Z}_3$ acts locally as the second action in \eq{trolo}. The Hodge numbers are $h_{1,1}=12$, split as three corresponding to Kahler parameters of the ambient torus and the other 9 being blow-up moduli of the singularities. When the singularity is blown up, we end up instead with a smooth geometry which is  the total space of a line bundle $\mathcal{L}$ over $\mathbb{P}^2$ with Chern class $-3$ times the canonical class of $\mathbb{P}^2$ (see e.g. \cite{Ganor:2002ae,GrootNibbelink:2007lua,GrootNibbelink:2007yig} and \cite{Junghans:2023yue} for a recent analysis of the blown-up DGKT solution). 

Finally, we also have $h_{2,1}=0$, which means this is a rigid CY orbifold. We now need to include the orientifold action. The geometric part $\iota_3$ of the orientifold involution acts as\footnote{The orientifold involution presented in the original paper \cite{DeWolfe:2005uu} had a minus sign in \eqref{ori9}. However, as discussed in \cite{Junghans:2023yue}, that orientifold involution is inconsistent with the orbifold action \eqref{trolo}; as explained in detail in that reference, it would yield a self-intersecting orientifold even after blowing up the orbifold singularities. This is impossible, as the fixed-locus of a differentiable involution in a smooth manifold cannot self-intersect (see e.g. Theorem 1.10.15 of \cite{klingenberg1995riemannian}. This different orientifold involution does not change the main results of the DGKT moduli stabilization, but it may affect the analysis of the twisted sectors.}
\begin{equation} z_i\,\rightarrow \bar{z}_i,\label{ori9}\end{equation}
which leads to an $O6$ plane wrapped on the $\text{Im}{z_i}=0$ locus (and orbifold images of it). 
Interestingly, the orientifold locus is not a smooth manifold, and it self-intersects at the orbifold fixed points where the orientifold locus looks locally like the union of three orientifold planes, as
\begin{equation} \{\text{Im}{z_i}=0\}\cup \{\text{Im}{\alpha^2\, z_i}=0\}\cup \{\text{Im}{\alpha^4\, z_i}=0\}\end{equation}. The blow-up deformations to the Kahler class are odd under the orientifold action and so, as described above, they survive the orientifold. In the resolved smooth geometry, the orientifold wraps a smooth immersed submanifold without self-intersection. In particular, this submanifold intersects the singular $\mathbb{P}^1$ at the orbifold locus. 

In short, the DGKT benchmark is a $T^6/\mathbb{Z}_3\times \mathbb{Z}_3$ orientifold with no complex structure moduli and 12 Kahler moduli, all of which are stabilized via the general DGKT mechanism discussed above. The geometric picture at large $N$ is of a very large, weakly curved toroidal orbifold, threaded by very dilute $F_4$ fluxes (the Roman's mass is a 10d cosmological constant, and does not dilute in string units). Except perhaps at the singularities, the solution is expected to be very weakly warped and with a nearly constant axiodilaton gradient. Since the dilaton is very weak, the gravitational backreaction of the Roman's mass and the other fluxes is almost negligible. The stabilized value for the blow-up modes can be tuned away from the orbifold point by turning on $F_2$ and $F_4$ fluxes on the local $\mathbb{P}^1$  and $\mathbb{P}^2$ classes described above; as described in \cite{DeWolfe:2005uu}, this is enough to ensure stabilization at a non-zero volume (and remove any orientifold self-intersections). From the EFT point of view, nothing goes wrong if one stabilizes at the orbifold point; but as we will discuss in the next Subsection, there are some reasons to avoid them.
 
 \subsection{What is weird about the DGKT solution?}
 
 The DGKT solution is one of the few proposals for scale-separated AdS vacua in string theory, and the most concrete one by far. At the same time, the solution is not at the level of control that, say,  the IIB AdS$_5\times S^5$ vacuum \cite{Maldacena:1997re}. DGKT faces a number of largely unresolved issues, which have resulted in a lively literature on the topic, particularly in recent years (see \cite{Lust:2004ig,Villadoro:2005cu,Acharya:2006ne,Aldazabal:2007sn,Banks:2006hg,Behrndt:2004km,Behrndt:2004mj,Martucci:2005ht,Grana:2006kf,Caviezel:2008ik,Aharony:2010af,Blaback:2010sj,Douglas:2010rt,McOrist:2012yc,Saracco:2012wc,Saracco:2013aoa,Gautason:2015tig,Font:2019uva,Marchesano:2019hfb,Baines:2020dmu,Junghans:2020acz,Buratti:2020kda,Marchesano:2020qvg,Cribiori:2021djm,Marchesano:2022rpr,Quirant:2022fpn,Apers:2022tfm,Apers:2022vfp,Apers:2022zjx,Andriot:2023fss,Carrasco:2023hta,Junghans:2023yue}). We will now provide a brief account of these, as well as of other issues not usually mentioned in the literature and whose importance we feel is minor.
 
The basic problem of the DGKT construction, and the one that has received the most attention, is the lack of a fully explicit 10-dimensional embedding of the solution \cite{Coudarchet:2023mfs}. As described in the previous Subsection, the DGKT scenario works by first finding the dimensional reduction of massive IIA string theory on a Calabi-Yau, writing down the corresponding 4d $\mathcal{N}=1$ EFT including fluxes, and then finding an AdS solution to the corresponding four-dimensional equations of motion. The issue is that a solution of these 4d equations of motion need not uplift to a full solution of the 10-dimensional equations of motion. The corresponding 4d EFT has not, to our knowledge, been proven to be a consistent truncation of the ten-dimensional supergravity. This means that there may be instabilities of the solution that are invisible to the 4d EFT, because they involve KK modes or some other ingredient that is integrated out in the four-dimensional equations of motion. One may think that, since the 4d solution to the 4d EFT is supersymmetric, it is guaranteed to remain SUSY and stable in the ten-dimensional theory; however to show that a given solution is supersymmetric, one must solve \emph{all} F-term and D-term equations, not just the ones in a given consistent truncation. There could be something like a KK F-term that is not solved, and then the solution would fail to be supersymmetric in ten dimensions.

Because the only localized brane in the minimal DGKT construction is the $O6$ plane, this issue is often referred to as the problem of ``orientifold smearing'' in the literature \cite{Baines:2020dmu} (where the term refers to the fact that the 4d equations of motion amount to integrating the 10d ones over the internal space, so all localized sources are ``smeared'').  The idea here is that the other ingredients in the construction (metric and fluxes) are not localized and are more likely to be adequately captured by the 4d EFT. But the issues of control of the 10-dimensional background go beyond orientifolds: for instance, since there are fluxes, the metric of the internal space will not be Calabi-Yau, and in principle, it can deviate significantly in regions of large curvature or coupling, such as near the orientifold or orbifold fixed points. There is a program \cite{Saracco:2012wc,Junghans:2020acz,Marchesano:2020qvg} aimed at solving better and better approximations to the ten-dimensional equations of motion with localized sources, 
but still from a four-dimensional EFT point of view. The 4d smeared DGKT vacuum arises then as the solution at first order in the expansion parameter\footnote{Although the expansion parameter in \cite{Marchesano:2020qvg} is $g_s$, this is because the solutions in the internal space source nonconstant dilaton profiles, and the perturbation parameter is a series expansion in dilaton inhomogeneities. These analyses do  \emph{not} include ten-dimensional $g_s$ corrections to the 10d action, which are largely not known.} $g_s$. In successive approximations, the metric is generalized from Calabi-Yau to $SU(3)\times SU(3)$ structure. So far, this program has not found any pathology, supporting the 10d uplift of the DGKT scenario, although it has only been carried out to second order\footnote{Potential issues related to having intersecting orientifolds could only become manifest at higher order in this expansion, since at first order the interaction between sources is negligible. However, going beyond second order on $g_s$ would require considering also $\alpha'$ corrections to the 10d action due to the relative scaling between the string dilaton and the overall volume in DGKT, which are not fully known.} in $g_s$.

A related issue that is often discussed is that, since the Calabi-Yau metric is warped, the estimate of the KK scale may be different from the naive one, which assumes an unwarped CY metric. Specifically, the estimate of the KK scale in \eq{estimates} comes from reading off the value of the Kahler moduli in the 4d solution, and then translate them to the size of the parent $T^6$ assuming a Ricci-flat metric. But this dictionary may change if the metric deviates from this.  This happens in a known example of flux compactification \cite{Font:2019uva}, where the additional warping was enough to correct the KK scale significantly, leading to KK and AdS curvature scaling in the same way. We believe this is unlikely to happen at least in the benchmark DGKT model, whose essential ingredient is a very large internal space with very low warping. 

Another concern that is often raised is the physics of orientifold loci, and in particular, of self-intersecting orientifolds \cite{Baines:2020dmu}. In massless type IIA, an $O6^-$ plane has a diverging dilaton at its core, and similar behavior may persist for $O6$ planes in massive IIA (although see \cite{Cremonesi:2015bld,Cordova:2019cvf}; there are several proposed massive IIA $O6$ solutions, and some have a diverging dilaton core, some do not). This divergence seems to become worse for intersecting $O6$ planes, such as the ones that appear in the benchmark DGKT model. The concern here is that, since the dilaton is locally very strong, the physics is actually out of control near the orientifold planes, and this could yield e.g. an instability mode that destroys the solution. This physics could in principle be stringy in origin, such as the localized spacetime-eating tachyons of \cite{Adams:2001sv}. 

However, $g_s$ blows up at the core of an $O6^-$ plane even in flat space, and this does not mean that we lose control of the configuration, which can be easily analyzed from the worldsheet. This is true even in massive IIA; we have explicit solutions involving $O6$ planes, where the orientifold locus behaves completely nicely \cite{Apruzzi:2017nck}. The key point is that the region where $g_s\gtrsim 1$ around an orientifold is smaller than one string length; that means that stringy probes, which have a resolution of order $l_s$, cannot see the singular core. This is a feature of all orientifold and D-brane solutions, and explains why they can be analyzed in the worldsheet\footnote{This does not mean that the worldsheet captures all degrees of freedom; we will see an example in the next paragraph.}; it is manifestly absent in e.g. the NS5 brane, whose classical solution has an infinite tube, of many string lengths in diameter, where the dilaton grows without bound.

What about a self-intersecting orientifold such as the one appearing above? We can certainly study perturbative orientifolds of orbifolds such as the one described in the benchmark DGKT model, with one small provision: the orbifold that can be described in worldsheet terms involves turning on a certain B-field in the configuration, which corresponds to tuning a Kahler modulus to a value where the worldsheet CFT becomes singular \cite{Douglas:2000qw}. Turning on this deformation of the singularity gives a mass to the tensionless strings arising from $D4$ branes wrapped on the collapsed cycle. If this  B-field is not turned on, then the orbifold has a certain localized 4d $\mathcal{N}=2$ SCFT living in its worldvolume, at least when the Roman's mass is not turned on (see Appendix \ref{app:A}). We have not studied the case when the Roman's mass is turned on (which typically amounts to Chern-Simons-like deformations of the SCFT \cite{Gaiotto:2009yz}); our point is that, in any case, we have techniques to determine these local degrees of freedom and they will usually be some local QFT. This may alter the set of degrees of freedom at low energies, but does not seem to affect significantly the DGKT construction or source any instability. Finally, one can desingularize the orbifold via explicit blow-up if so desired, as described in the previous Subsection, and then the orientifold is not self-intersecting anymore, so the above questions may be ignored.

Both some of the nicest and weirdest features of the DGKT solution come from thinking about its holographic dual. Assuming DGKT exists, as well as the AdS/CFT correspondence (see e.g. \cite{Nastase:2007kj} for an introduction), there must be a 3d $\mathcal{N}=1$ large $N$ family of dual field theories\footnote{With the usual notation, a 3d $\mathcal{N}=1$ SCFT has two ordinary and two ``conformal'' supercharges. These four supercharges are arranged as a single 4d $\mathcal{N}=1$ multiplet in the gravity dual. }. The DGKT vacuum was studied from this point of view in \cite{Aharony:2008wz}. SCFT's often come with continuous R-symmetries, whose holographic dual in all known examples are KK photons. The BPS operators charged under the R-symmetry are constrained to have low dimension by the BPS bound, dual to low masses in the bulk, of order $\ell_{\text{AdS}}^{-1}$. Since in all known examples there is an infinite tower of BPS states, there is an extra dimension of AdS size, and the AdS vacuum is not scale separated. The connection between continuous R-symmetries, existence of BPS states, and absence of scale separation is well-known \cite{Alday:2019qrf,Cribiori:2022trc}; in \cite{Montero:2022ghl} an argument against extreme scale separation was given which does not rely on BPS states, but only on the consistency of the flat-space limit of these theories. Another argument based on having a bona fide 't Hooft limit was recently given in \cite{Calderon-Infante:2024oed}.

The simplest loophole to these considerations is to look for an SCFT with so little supersymmetry that there is no continuous R-symmetry. As can be seen from the classification of superconformal algebras, the only possibilities are 3d $\mathcal{N}=1$ or 2d $(1,0)$ theories. Therefore, DGKT nicely has just the right amount of supersymmetry to achieve scale separation.  It is also worth noting that DGKT is actually consistent with the mild version of the AdS Distance Conjecture \cite{Lust:2019zwm}, since the Kaluza-Klein tower still becomes light in the flat space limit; it simply does so at a smaller rate than the AdS scale. Therefore, it is only inconsistent with the strong version of this conjecture, which fixes the rate such that no scale separation is possible in SUSY setups (see \cite{Buratti:2020kda} for a possible explanation in terms of discrete three-form symmetries).

The central charge of the 3d $\mathcal{N}=1$ large $N$ family of SCFT's dual to DGKT can be read off from \eq{estimates}. It scales with the $F_4$ flux as 
\begin{equation} c\sim N^{\frac92}.\label{crazyc}\end{equation}
In our opinion, this is one of the most peculiar features of the DGKT vacuum. Such a rapid scaling with $N$ is unheard of in other controlled AdS flux compactifications with holographic dual\footnote{Taking $N$ to be the ``total number of branes'' used to construct the geometry, or equivalently, the classical dimension of the moduli space of the dual field theory. If one focuses on e.g. one single type of brane charge, any scaling with $N$ is possible, e.g. take the D1-D5 SCFT with $N_1\sim N_2^k$ for any $k$, then $c\sim N_2^{k+1}$, but $c$ only scales at most quadratically on the total number of branes $N=N_1+N_2$. }. The usual picture in holography is that, if one has a large $N$ family of flux compactifications with some particular flux(es) $F\sim N$ turned on, the dual field theory can be constructed as the worldvolume QFT of $N$ branes probing a singularity, where the brane is precisely the one that can turn on the flux $F$. The original flux compactification is then recovered as the near-horizon geometry of the brane stack. For instance, for the IIB AdS$_5\times S^5$ solution, the field theory is generated by a stack of $N$ D3 branes, and the central charge goes as $N^2$ as is familiar from the worldvolume gauge theory. For M-theory on AdS$_4\times S^7$, $c\sim N^{3/2}$, which comes from the dual ABJM theory \cite{Aharony:2008ug}. We have examples of five-dimensional SCFT's whose central charge grows as $N^{5/3}$ \cite{Guarino:2015jca}, and M-theory on AdS$_7\times S^4$ (dual to the $(2,0)$ wordvolume QFT of a stack of $N$ M5-branes), gives us a central charge of $N^3$, which is the largest value that has been found in holographic examples. To our knowledge, nowhere in the brane zoo a brane which gives \eq{crazyc} can be found. Furthermore, the large $N$ flux in the DGKT scenario is an $F_4$ flux, so the dual geometry should be constructed simply with a stack of $D4$ branes. It is conceivable that D4 branes probing a very exotic, yet to be discovered singularity yields \eq{crazyc} (perhaps via multi-pronged strings or branes becoming tensionless \cite{Banks:2006hg}), but this has not been constructed yet.

Another intriguing discussion in the recent literature concerns the appearance of integer conformal dimensions. As first noted in \cite{Conlon:2021cjk}, the masses predicted by the 4d $\mathcal{N}=1$ EFT (and which become exact in the infinite $N$ limit where the ten-dimensional dilaton vanishes) only take a particular set of values, which are integral when expressed in units of $\ell_{AdS}$. Furthermore, these values are fine-tuned so that the conformal dimensions of the dual scalars, related to the masses by the quadratic equation
\begin{equation}\Delta(\Delta-3)= m^2\ell_{AdS}^2\end{equation}
are themselves also integral. This behavior is highly non-generic, and has been observed in a range of models \cite{Apers:2022zjx,Apers:2022tfm,Quirant:2022fpn,Apers:2022vfp,Andriot:2023fss}, both scale-separated and non-scale separated. Possible explanations include the emergence of Galilean-like symmetries in the large $N$ limit \cite{Apers:2022vfp} or some sort of enhancement to (a truncation of) an $\mathcal{N}=2$ field theory in the infinite $N$ limit (in an $\mathcal{N}=2$ theory, integral conformal dimensions are natural for BPS operators).  Although the issue is very interesting and there is no final word on it yet, at present we do not see that it has any implications for the consistency of the DGKT setup. 

Another exotic feature was recently pointed out by \cite{Bobev:2023dwx}. The light scale-separated DGKT spectrum seems to yield exotic logarithmic corrections to the black hole entropy (when taking the EFT cutoff to be at the KK scale), which from the CFT perspective would depend on continuous parameters; a very exotic feature never seen in known holographic duals and that it is not expected to appear in CFT's obtained from RG flow of an ordinary gauge theory.

Finally, if the 4d $\mathcal{N}=1$ EFT analysis is to be trusted, there are no marginal or relevant deformations in the DGKT geometry, and hence, it constitutes an example of a ``dead-end CFT'', which can be rigorously shown to not exist at least perturbatively \cite{Nakayama:2015bwa}. Again, all known explicit top-down examples of holography in $d>2$ admit marginal deformations, so this is yet another alien feature of the holographic dual to DGKT.

To sum up, all the above points may be more or less concerning according to taste, but none of them is an immediate killer. The rest of the paper is devoted to studying in some detail one more point to the list, which was briefly alluded to in \cite{Aharony:2008wz}: The DGKT setup has so little supersymmetry, that quantum corrections to the moduli space might lift it completely. To understand this, we now turn to a general study of the moduli space of holographic QFT's. 
 
 \section{CFT moduli space of 4d \texorpdfstring{$N=1$}{N=1} AdS vacua} \label{sec:s3}
 We will begin with a general discussion about moduli spaces in quantum field theory and holography, which then will be particularized to flux compactifications and the DGKT solution.
 
 \subsection{General properties of moduli spaces in holographic CFT's}\label{sec:brbr}
 
\begin{figure}[htb!]
\begin{center}
\includegraphics[scale = 0.75]{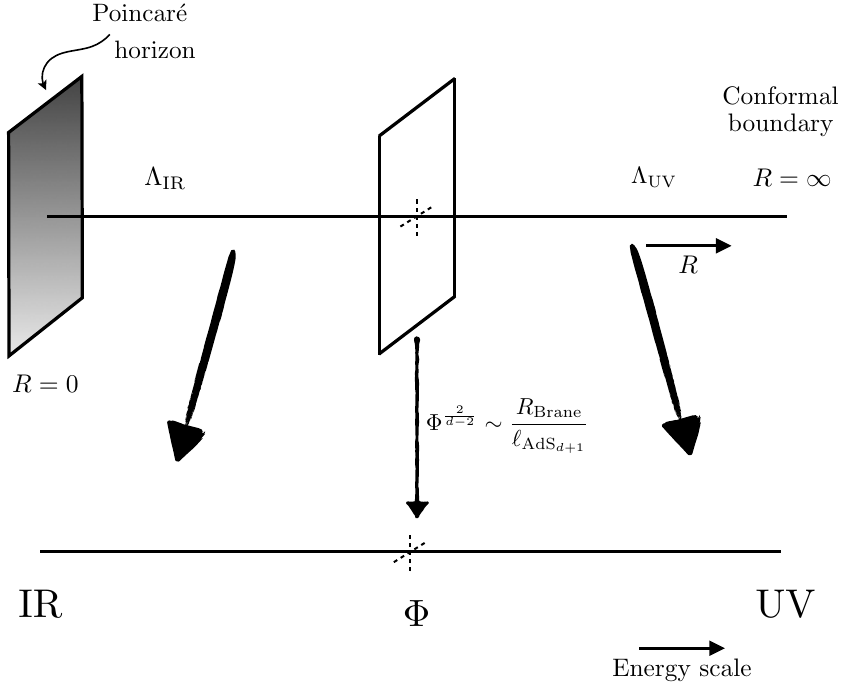}
\caption{Diagram illustrating the holographic correspondence between a brane parallel to the Poincar\'{e} horizon in AdS (top) and the dual CFT description (bottom). The brane is a domain wall separating two regions of generally different cosmological constant. Dually, the position of the brane in the AdS Poincar\'{e} coordinate becomes the vev of some field $\Phi$. The region close to the conformal boundary remains identical to the original CFT vacuum, and corresponds to the deep UV in the field theory side, where the scalar vev may be neglected. Contrarily, the region close to the Poincar\'{e} horizon, which is also locally AdS, corresponds to the deep IR, where the local physics is captured by a different CFT (appearing in the IR when the vev for $\Phi$ is turned on). Whether there is a scale-invariant potential for $\Phi$ in the bottom picture maps to the force felt by the brane in the top picture; the case of no scalar potential for $\Phi$ corresponds to the brane being BPS.}
\label{setup-fields}
\end{center}
\end{figure}

Quantum field theories often have nontrivial vacuum manifolds, parametrized by the vacuum expectation values of one or more fields $\Phi$. In general, there is an arbitrary scalar potential $V(\Phi)$ depending on the vacuum expectation values of these fields. In the most general case, $V(\Phi)$ is completely unconstrained. However, the scalar potential can be constrained by the presence of additional symmetries. For instance, with supersymmetry, $V$ must be non-negative, and it is often the case that $V$ exactly vanishes for some special directions, which parametrize the so-called moduli space. The study of the full moduli space of vacua has been very fruitful in revealing information about the QFT and those special points in moduli space where superconformal symmetry appears, dubbed SCFT points. This includes holographic SCFT's, in which case, the moduli space has a natural picture in the gravitational dual (see e.g. \cite{Maldacena:1997re,Papadopoulos:2000hb,Aharony:2008wz}) : it is simply the configuration space of systems of BPS branes in the AdS theory in Poincar\'{e} patch. Do not confuse this with the moduli space of the gravitational bulk dual, which is dual to the space of marginal deformations of the SCFT. Contrarily, the moduli space coordinates of the SCFT correspond to the locations of BPS branes in the bulk AdS Poincar\'{e} coordinate. In the holographic dual, this is in turn represented by having one or more BPS branes in the bulk, at a constant Poincar\'{e} radius coordinate (see Figure \ref{setup-fields}).  The existence of the moduli space, i.e. that there are vevs that one can turn on without increasing the energy of the system, is mapped to the fact that the dual BPS branes satisfy a BPS no-force condition\footnote{The BPS condition is much more intricate in AdS than in flat space, and there are many possible BPS brane configurations (see \cite{Skenderis:2002vf,Skenderis:2002ps,Skenderis:2003da}). The planar configurations we discuss in this Section are the only ones that preserve the Poincar\'{e} symmetry of the dual CFT, as must be the case for the moduli space.}. As a result, one can move them in the Poincar\'{e} direction at zero cost in energy.

Our goal will be to understand the DGKT geometry by understanding the moduli space of the dual field theory. We will begin by reviewing in some detail the constraints that supersymmetry and conformal symmetry place on the scalar potential $V(\Phi)$ of a quantum field theory, and understanding the holographic dual of these statements in terms of D-brane physics.  To this effect, consider a $d>2$-dimensional QFT with a space of vacua parametrized by a coordinate $\Phi$ and suppose that conformal symmetry is restored at a particular point, say $\Phi=0$.  Since at $\Phi=0$ we have a conformal field theory and the conformal symmetry is unbroken, but the scalar potential must respect this symmetry even when $\Phi\neq0$, assuming that the conformal symmetry is spontaneously but not explicitly broken. In particular,   $V(\Phi)$ cannot depend on any scales other than $\Phi$. As a result, choosing $\Phi$ so that it is canonically normalized (so that it has dimension $(d-2)/2$),  the potential is fixed by dimensional analysis to (see e.g.\cite{Seiberg:1999xz})
\begin{equation} V(\Phi)= \lambda\, \Phi^{\frac{2d}{d-2}},\label{brascapo}\end{equation}
where $\lambda$ is a dimensionless coupling. If there is more than one scalar field, $\lambda$ may depend on ratios of vevs. But upon a homogeneous rescaling of all the vevs, $V$ must be homogeneous\footnote{The case $d=2$ must be treated separately, but there is also universality; the effective theory of $\Phi$ becomes Liouville theory in this case \cite{Seiberg:1999xz}. } of degree $2d/(d-2)$. 

As mentioned before, in a holographic CFT, turning on a vev $\Phi$ corresponds to putting a codimension-one brane, or domain wall, in the bulk. In Poincar\'{e}-AdS coordinates,
\begin{equation} ds_{\text{AdS}_{d+1}}^2= \frac{dR^2}{R^2}+ \frac{R^2}{\ell_{\text{AdS}_{d+1}}^2}\left( -dX_0^2+ \sum_{i=1}^{d-1} dX_d^2\right),\label{poads}\end{equation}
the brane spans the $T$ and $X_i$ directions, while sitting at a concrete value of the $R$ coordinate. This preserves exactly the same isometries of AdS than a scalar vev does in the dual field theory; furthermore, for large $R$, far away from the brane, the configuration looks like the AdS vacuum. The cosmological constant typically jumps across the brane, so for small $R$, close to the Poincar\'{e} horizon at $R=0$, the physics looks like that of a different conformal field theory. Using the UV/IR correspondence \cite{Peet:1998wn}, this matches CFT expectations: At very low energies, below $\Phi$, the QFT flows to some IR CFT, while at very high energies, the vev $\Phi$ can be ignored and physics is described by the original CFT (see Figure \ref{setup-fields}). 

What is the meaning of $\lambda$ from the bulk point of view? The scalar potential is just the energy of the configuration, so we must compute the vacuum energy of the brane worldvolume. The force exerted on the codimension-1 brane is governed by two competing effects. On one hand, there's the kinetic term, of the form
\begin{equation} T\int dV_{\text{Ind}},\end{equation}
where $dV_{\text{Ind}}$ is the induced metric on the brane worldvolume. For the brane configuration under consideration,
\begin{equation} dV_{\text{Ind}}= \left(\frac{R}{\ell_{\text{AdS}_{d+1}}}\right)^{d}\, dX_0\wedge \dots\wedge dX_{d-1}\label{wepopo}\end{equation} On the other hand, a codimension-one brane couples to a $C_{d}$-dimensional potential, as
\begin{equation} -\tilde{q} \int C_{d}.\end{equation}
It is compatible with AdS isometries (and it happens in the cases of interest in this paper) that the AdS geometry is permeated by a top-form field strength,
\begin{equation} F_d=dC_{d}= \kappa\, dV_{\text{AdS}_{d+1}}\quad\Rightarrow\quad C_{d-1}= \frac{\kappa}{d}\, dV_{\text{Ind}} \end{equation}
where in the last equality we have used \eq{wepopo} and the explicit expression for the volume form of the metric \eq{poads}. Using this, the brane vacuum energy turns out to be
\begin{equation}  T\int dV_{\text{Ind}} -\tilde{q} \int C_{d}= \left( T- \frac{\tilde{q} \kappa}{d}\right)\int dV_{\text{Ind}}= (T-q)\left(\frac{R}{\ell_{\text{AdS}_{d+1}}}\right)^{d} \int \, dX_0\wedge\dots\wedge dX_{d-1},\label{bulkbrane}\end{equation}
where in the last equality we have defined 
\begin{equation}q\equiv\frac{\tilde{q} \kappa}{d}. \end{equation}
Equation \eq{bulkbrane} has the form of a dual field theory scalar potential \eq{brascapo}, with the identifications
\begin{equation} \Phi^{\frac{2}{d-2}}\sim\left(\frac{R}{\ell_{\text{AdS}_{d+1}}}\right),\quad \lambda \sim T-q.\label{awoo}\end{equation}
As a cross-check, $ \Phi^{\frac{d}{d-2}}$ has units of energy under scale transformations. These correspond to rescalings of the radial coordinate $R$ in Poincar\'{e} patch so, indeed, \eq{awoo} matches the symmetries of the system. What we learned is that the quantity $\lambda$ controls the difference  between the mass/tension of a brane and the force it feels in the ambient AdS background. The conformal point corresponds to tuning the D-brane position at $R=0$, where it disappears behind the Poincar\'{e} horizon, and we recover empty AdS space. 

Consider now the special case where the theory is supersymmetric, and $\Phi$ is a direction in the moduli space. This forces $\lambda=0$, so in the bulk dual of the field theory, we have 
\begin{equation} T=q.\label{BPSc}\end{equation}
The force felt by the branes exactly cancels against the gravitational attraction of the AdS well: The branes are BPS. They preserve some supersymmetry, which in turn matches the SUSY preserved in the dual field theory at $\Phi \neq 0$, and they feel no force because of that. In other words, the dual statement of the existence of a CFT moduli space is that a brane saturates the BPS condition \eq{BPSc}. Furthermore, the BPS condition \eq{BPSc} is an absolute lower bound on the tension of any brane in the theory, which must satisfy 
\begin{equation} T\geq q\label{bpsco}\end{equation}
with equality if and only if the brane is BPS. The dual statement is that $\lambda\geq0$, which is a particular instance of the fact that supersymmetric quantum field theories have a non-negative scalar potential \cite{freedman2012supergravity}.

As studied in detail in \cite{Ooguri:2016pdq} (see also \cite{Seiberg:1999xz,Maldacena:1998uz}), the BPS condition \eq{BPSc} and bound \eq{bpsco} are intimately related to the Weak Gravity Conjecture \cite{ArkaniHamed:2006dz}. In its version for membranes, the WGC demands that there is a membrane state satisfying  
\begin{equation} T \leq q \label{wgcmem}.\end{equation}
In a supersymmetric theory, because of the BPS bound \eq{bpsco}, \eq{wgcmem} can only be satisfied when $T=q$, i.e. when there is an exactly BPS membrane\footnote{The main point of \cite{Ooguri:2016pdq} is the stronger claim that \eq{wgcmem} can be saturated if and only if the AdS vacuum is supersymmetric. Since $T<q$ corresponds to $\lambda<0$, and the scalar potential is unbounded below, there can be no unitary CFT in this case, and the dual AdS vacuum is unstable to bubble nucleation. Thus, \cite{Ooguri:2016pdq} concluded that no nonsupersymmetric AdS can have a unitary holographic dual. Although this is extremely interesting and has led to a large following in the literature, in this paper we will only assume the slightly weaker version of WGC for membranes discussed in the main text.}. Therefore, the WGC for membranes can be recast, in the dual CFT language, as the statement that

\begin{center} \textbf{The dual CFT to a holographic, stable AdS vacuum supported by fluxes\\ must have an exact moduli space of deformations. }\end{center}

 As a simple example of all this, consider a stack of $N$ $D3$ branes in $\mathbb{R}^{10}$, whose worldvolume theory is $SU(N)$ $\mathcal{N}=4$ SYM and whose holographic dual is the famous AdS$_5\times S^5$ solution. One can give a vev to any of the $SU(N)$-valued scalars of the theory; the physical interpretation for a diagonal vev is that one or more of the $D3$ branes are separated from the main stack \cite{Maldacena:1997re}. These branes are BPS, satisfying \eq{BPSc} exactly. Each of the D3 branes carries its own worldvolume field theory in the bulk. For instance, consider the Higgsing pattern in $\mathcal{N}=4$ SYM
\begin{equation}SU(N)\,\rightarrow SU(N-k)\times SU(k_1)\times SU(k_1)\ldots\label{hjyt}\end{equation}
for a configuration where $k$ branes have been moved and they form stacks of $k_1,k_2,\ldots$ in the bulk. The right hand side of \eq{hjyt} also describes the worldvolume gauge theories in each of the stacks. Since the branes are domain walls for the $F_5$ flux that supports the solution, the AdS space in between the two stacks does not have the same cosmological constant, which jumps when crossing a brane stack. Only near the AdS boundary we recover the cosmological constant corresponding to a stack of $N$ branes; as we decrease the radial direction, the absolute value of the cosmological constant goes up, and near the Poincar\'{e} horizon we recover the value corresponding to a stack of $N-k$ $D3$ branes. This matches the usual UV/IR correspondence in holography; In the UV (near the boundary), the vevs we turned on are negligible and the solution looks like $SU(N)$ $\mathcal{N}=4$ SYM, while in the IR, the vevs have completely higgsed the gauge group to $SU(N-k)$, as explained above.  We can play a similar game in DGKT, since it is also a large $N$ flux compactification. The corresponding domain walls and moduli space were studied in some detail in \cite{Aharony:2008wz}. The four-dimensional domain walls are $D4$ branes wrapped in the cycle dual to the one where the $F_4$ flux is switched on.

In the absence of supersymmetry, the WGC is expected to be fullfiled by membranes satisfying \eqref{wgcmem} but not saturating it, since nothing protects the worldvolume theory from receiving quantum corrections that make $\lambda\neq 0$. This is what led \cite{Ooguri:2016pdq} to conjecture that non-SUSY AdS vacua supported by fluxes can be at best metastable, since a membrane with $T<q$ feels a repulsive force towards the Poincare horizon \cite{Maldacena:1998uz}, yielding a non-perturbative instability that discharges the flux vacuum. Contrarily, membranes with $T>q$ feel an attractive force towards the Poincare horizon, which is consistent with stability of the SUSY vacuum. In the next Section we will see that 4d $\mathcal{N}=1$ vacua are similar to non-SUSY vacua in the sense that there is typically too little supersymmetry to protect the field theory from quantum corrections that generate a non-zero $\lambda$. However, unlike in the non-SUSY case, the BPS bound \eqref{bpsco} makes impossible to get membranes satisfying the WGC with $T<q$. This presents an intriguing conundrum, which we explore further in this paper.

\subsection{Moduli spaces and quantum corrections}\label{subs:modq}

From a field theory perspective, the existence of a moduli space and its stability against corrections is intimately tied to extended supersymmetry. For instance, in the $\mathcal{N}=4$ SYM example that we described in the preceding paragraph, the moduli space cannot receive quantum corrections, and in particular, none of the moduli can acquire a mass. In situations with less supersymmetry, such as a 4d $\mathcal{N}=2$ or 3d $\mathcal{N}=4$ SCFT, some moduli directions can become massive, but only at special points in moduli space where the right kind of multiplets can be gapped out together. For instance, a vector multiplet can only become massive if there is a charged hypermultiplet with which it can pair via a Higgs mechanism (as it happens in a conifold transition), but in absence of this, it cannot get a mass whatsoever. In the 4d $\mathcal{N}=2$ case, the theory of the moduli space (Seiberg-Witten theory) gains much of its power due to the fact that the supersymmetry requires the superpotential to be a holomorphic function of the moduli, which is therefore severly constrained.

The situation is drastically different for 3d $\mathcal{N}=1$ theories, such as the putative dual to DGKT. In 3d $\mathcal{N}=1$ theories, the supercharge is a single Majorana spinor, and consequently, there is a single real superspace coordinate $\theta^\alpha$. As a result, when constructing a superspace action\footnote{See the first chapter of \cite{Gates:1983nr} for a detailed introduction to 3d $\mathcal{N}=1$ superspace.}, there is no $d^2\theta d^2\bar{\theta}$ term, and the whole superspace action is written in terms of a single superpotential $\mathcal{W}$:
\begin{equation}S=\int d^2\theta\, \mathcal{W}.\end{equation}
$\mathcal{W}$ contains both kinetic terms, potentials, and interactions for all fields. Since $\theta^\alpha$ is a real spinor parameter, there is no holomorphicity, and therefore, \emph{the moduli space is not protected against quantum corrections}. This means that a superfield $R$, such as the moduli space coordinate considered above, can receive a potential via loop effects, even if  $\partial_R\mathcal{W}=0$ classically (as explained in \cite{Choi:2018ohn}, the scalar potential can only appear a two loops). The field theory literature contains several examples of this phenomenon, even in simple theories \cite{Bashmakov:2018wts,Gaiotto:2018yjh,Choi:2018ohn}. 

The obvious concern is then whether the moduli space of the 3d $\mathcal{N}=1$ field theory dual to DGKT is stable against quantum corrections. If it isn't, and all moduli space directions are lifted, then we will be forced to conclude that the WGC for membranes does not hold in the DGKT solution. More concretely, we could consider moving a single D4 brane at fixed Poincar\'{e} coordinate in the bulk. Its low-energy EFT is also a 3d $\mathcal{N}=1$ field theory, and the only noncompact scalar there is the position of the brane $R$. As we will see in Section \ref{sec:modquantum}, this direction classically has no potential, but it may be lifted by quantum effects. Understanding this is the main point of the paper. 

To sum up we face a trichotomy. Either:
\begin{enumerate}
\item The holographic dual of the DGKT solution has exactly massless directions, in spite of the fact it is a 3d $\mathcal{N}=1$ field theory where these are generically lifted,
\item or the WGC for membranes is wrong in AdS vacua,
\item or there is some pathology in the DKGT solution that destroys it or at least renders it non-supersymmetric.\end{enumerate}

The rest of the paper is devoted to exploring this conundrum. We spend the remainder of this Section providing general arguments against the first possibility, which is then excluded by explicit calculation in Section \ref{sec:modquantum}. This only leaves either the second or the third possibility as possible outcomes. We don't know which one is right, but offer some speculations in the Conclusions.

 \subsection{3d \texorpdfstring{$\mathcal{N}=1$}{N=1} moduli spaces and parity symmetries}
 
The reader might be bothered by the discussion in the previous Subsection, which relies on supersymmetry alone, and does not rely on scale separation or other detailed DGKT features at all. Taken at face value, it would seem to suggest that there are no holographic 3d $\mathcal{N}=1$ field theories (at least with Einstein duals) at all. But there in fact are many examples of such theories; one particularly simple construction involves a stack of $N$ M2 branes probing a codimension-eight singularity in M theory, which is a cone over a weakly $G_2$ manifold. The geometry breaks the 32 M-theory supercharges to just 2, which is the same supersymmetry preserved by the M2 branes, yielding a correspondence
\begin{equation}\text{N M2's on singularity}\,\leftrightarrow\, \text{M on AdS$_4\times WG_2$}.\end{equation}
The right hand side is a 4d $\mathcal{N}=1$ flux compactification of M-theory. Explicit examples were first constructed as M-theory orientifolds of Calabi-Yau fourfolds \cite{Forcella:2009jj};  see \cite{Amariti:2014ewa,Franco:2021ixh} for related work. The dual field theory is a particular orientifold of an ABJM-like quiver. 

This setup has none of the funny properties of DGKT. It is not scale-separated (in the explicit examples discussed so far), its central charge scales as a reasonable $N^{\frac32}$, and one can check explicitly that the $M2$ branes are exactly BPS and there are no corrections to the field theory superpotential. This is most easily done in the Calabi-Yau orientifold examples of \cite{Forcella:2009jj}: Without the orientifold, the setup actually describes a configuration with higher, 3d $\mathcal{N}=2$ supersymmetry, and one just has to focus on orientifold-invariant configurations. There are no higher-derivative corrections, and one can explicitly check that there are no non-perturbative effects (such as Euclidean M5 branes) that could contribute\footnote{This setup is like the ones discussed in \cite{Palti:2020qlc}, which inherits some features of a higher supersymmetric parent model.
}.

It turns out that this protection against corrections is not accidental at all, and there is a physical reason for it: The existence of a parity symmetry. Although 3d $\mathcal{N}=1$ is not enough SUSY to guarantee protection against quantum corrections,  3d $\mathcal{N}=1$  in combination with a parity symmetry is enough to do so \cite{Gaiotto:2018yjh}. We will describe this mechanism briefly, referring the interested reader to \cite{Gaiotto:2018yjh} for an extended discussion.

A parity transformation is merely a flip of one of the spatial coordinates, say $x_1\,\rightarrow -x_1$. Since supersymmetric theories contain fermions, we must define the action of parity on fermions, in a manner consistent with the Lorentz algebra. One possibility is to set
\begin{equation} \psi(x,y,t)\,\rightarrow \gamma^1\, \psi(-x,y,t),\label{gammar}\end{equation}
where $\psi$ is a fermion field and $\gamma^1$ is a gamma matrix satisfying the usual anticommutation relations $\{\gamma^\mu,\gamma^\nu\}=2\eta^{\mu\nu}$,
and $\eta^{\mu\mu}\equiv\text{diag}(-1,1,1,1)$. Defined in this way, the parity transformation squares to the identity, $\mathrm{P}^2=1$ both in fermions and bosons. However, it is also possible to multiply the $\gamma^1$ by $i$ in \eq{gammar}, and the resulting parity transformation satisfies instead $\mathrm{P}^2=(-1)^F$. These two possibilities are called Pin$^+$ (if parity squares to +1) and Pin$^-$ (if parity squares to $(-1)^F$). Both are realized in different physical systems \cite{Witten:2015aba}, but with 3d $\mathcal{N}=1$ supersymmetry, only Pin$^+$ is possible \cite{Berg:2000ne,Montero:2020icj}. This is because in 3d $\mathcal{N}=1$, the supercharge is real (transforms as a Majorana spinor), and so we are not allowed to multiply by $i$ in \eq{gammar} (that would make the spinor complex, doubling the number of degrees of freedom). A basis of real, Lorentzian gamma matrices is given by \cite{polchinski1998string}
\begin{equation}\gamma^0=i\sigma^2=\left(\begin{array}{cc}0&1\\-1&0\end{array}\right),\quad\gamma^1=\sigma^1=\left(\begin{array}{cc}0&1\\1&0\end{array}\right),\quad \gamma^2=\sigma^3=\left(\begin{array}{cc}1&0\\0&-1\end{array}\right). \label{gammabasis}\end{equation}

In a parity invariant theory, the Lagrangian must be parity invariant. The superspace coordinates $\theta^\alpha$ transform as in \eq{gammar}, and the superspace measure, which is
\begin{equation}d^2\theta= (\gamma^0)_{\alpha\beta} d\theta^\alpha d\theta^\beta= d\theta^T \gamma^t d\theta \end{equation}
transforms as (say, under a reflection of the $x$ axis),
\begin{equation} d^2\theta\,\rightarrow  d\theta^T(\gamma^1)^T \gamma^0  (\gamma^1)\, d\theta= -d^2\theta,\end{equation}
where we have used \eq{gammabasis}.
In other words, the superspace measure is odd under parity. For the Lagrangian to be invariant, this means that the superpotential $\mathcal{W}$ must also be parity odd. A simple argument similar to the one above shows that standard kinetic terms for any field are odd. However, when considering terms with no derivatives, one learns that a parity-even scalar $R$ cannot appear in the superpotential, simply because any function of $R$ will be parity-even. Furthermore, a parity-odd scalar $Y$ can have superpotential terms, but they must be odd functions. So for instance, a $Y^3$ interaction is allowed, but a mass term $Y^2$ is not. 

There are concrete examples of controlled 3d $\mathcal{N}=1$ holographic vacua with an Einstein dual (so that the arguments that we have been making apply) where there is a moduli space protected by a parity symmetry\footnote{One example of a 3d $\mathcal{N}=1$ large $N$ holographic SCFT without parity symmetry is the $O(N)$ vector model, which has a quartic superpotential (see \cite{Moshe:2003xn,Aharony:2019mbc}). This theory is holographic, but not Einstein, and it does not have a moduli space. In Einstein holographic theories one expects a moduli space to always exist due to the WGC, as described in Subsection \ref{subs:modq}. We thank A. Sharon for explaining this example to us.}. In the M-theory vacua of \cite{Forcella:2009jj}, the parity symmetry is manifest in the bulk side: M theory has a Pin$^+$ symmetry \cite{Horava:1996ma}, under which the three-form $C_3$ is odd. Under a parity transformation in one of the AdS$_4$ directions, say of a direction orthogonal to the radial direction, the volume form of AdS$_4$ flips sign, so a  $G_4$ flux proportional to $dV_{\text{AdS}_4}$ remain invariant. The same argument holds in a configuration with an $M2$ brane in the bulk. In this case, the field  $R$ parametrizing the radial position of the $M2$ brane is even under the parity symmetry, ensuring that the condition $\partial_R \mathcal{W}=0$ holds even at the non-perturbative level. 

Having understood the existing 3d $\mathcal{N}=1$ vacua, the obvious question is whether the same mechanism is at place in DGKT. This is what we will discuss next.

\subsection{Discrete and parity symmetries in DGKT vacua}\label{sec:parDGKT}

We will now study the discrete and parity symmetries of type IIA strings, and their fate in the DGKT vacuum. The IIA GSO projection of type II strings leads to a chiral worldsheet theory \cite{polchinski1998string}, and because of this, the orientifold transformation $\Omega$, which merely reverses the orientation of the string worldsheet, is not a good symmetry of the theory. In fact, it is only a symmetry when combined with an odd number of reflections, such as the $\iota_3$ in Section \ref{sec:s2}. On the other hand, $(-1)^{F_L}\Omega$ is a good perturbative symmetry, used in the construction of reduced rank backgrounds with 16 supercharges \cite{Hellerman:2005ja}. This state of affairs contrasts with IIB, in which both $\Omega$ and $(-1)^{F_L}$ are good symmetries.  Instead, in IIA, only $\mathcal{C}=(-1)^{F_L}\Omega$ is a good internal symmetry of the worldsheet. Type IIA string theory is also nonchiral in spacetime, leading to a parity symmetry $\mathcal{P}$. The algebra generated by $\mathcal{P}$ and $\mathcal{C}$ is
\begin{equation} \mathcal{P}^2=\mathcal{C}^2=+1,\quad \mathcal{C}\, \mathcal{P}+(-1)^F \mathcal{P}\,\mathcal{C}=0.\label{alg0}\end{equation}
Since $\mathcal{P}$ is a parity symmetry that squares to $+1$ on fermions, it allows us to consider IIA string theory on non-orientable manifolds of the Pin$^+$ type; thanks to the algebra \eq{alg0}, the symmetry $\mathcal{C}\mathcal{P}$ squares to $-1$, allowing us to consider IIA string theory on non-orientable manifolds of Pin$^-$ type as well. These properties and the above algebra are most easily understood from the M-theory perspective. As discussed in the previous Subsection, M-theory has a celebrated parity symmetry of the Pin$^+$ type \cite{Horava:1996ma,Witten:2016cio,Freed:2019sco}, which also flips the sign of the $C_3$ field. This is required so that the triple Chern-Simons term
\begin{equation}\int C_3\wedge G_4\wedge G_4\end{equation}
is itself invariant. Regarding IIA as M-theory on $S^1$, the symmetry $\mathcal{P}$ corresponds simply to an M-theory parity acting on one of the non-compact coordinates, while $\mathcal{C}$ corresponds to a parity transformation along the $S^1$. This perspective also allows to efficiently deduct the transformations properties under $\mathcal{C}$ and $\mathcal{P}$ of the different IIA fields, without the need to go through a worldsheet analysis. These are listed in Table \ref{t1}, where a $\pm$ sign in the first row indicates that the field transforms in a standard or twisted representation for $\mathcal{P}$\footnote{Fields with a $+$ just change sign, under reflection, of their components along the reflected directions, while twisted fields, with a $-$ sign, change the sign of the components in all directions not being reflected. Think of the $\vec{E}$ and $\vec{B}$ fields of electromagnetism.}, while the sign in the second row directly describes the $\mathcal{C}$ action on the field.
\begin{table}[!htb]\begin{center}
\begin{tabular}{c|c|c|c|c|c|c|c}
Symmetry&$B_2$&$B_6$& $C_1$&$C_3$&$C_5$&$C_7$&$F_{10}$\\\hline
$\mathcal{P}$&-&+ &+&-&+&-&+\\
$\mathcal{C}$&+&-&-&-&+&+&+
\end{tabular}\end{center}
\caption{Action of the $\mathcal{C}$ and $\mathcal{P}$ symmetries on the different bosonic fields of IIA, as described in the main text. Notice that dual pairs of RR fields have opposite transformation properties under parity, which follows from the fact that the Hodge star operator $*$ anticommutes with the action of reflections on differential forms.}\label{t1}
\end{table}

Unlike the other fields, the transformation properties of $F_{10}$ (or its dual $F_0$, the Roman's mass $m$) cannot be argued directly from the M-theory description, since there is no M-theory lift of Roman's mass. Nevertheless, it can be obtained from consistency of Chern-Simons terms like
\begin{equation} F_0\, B_2 \wedge F_8\end{equation}
that are present in the massive IIA action. Since $B_2\wedge F_8$ is odd under  either $\mathcal{P}$ (as the top-form always picks the reflected direction) or $\mathcal{C}$, $m$ is odd under both.  

From the above table it is easy to see that the DGKT solution spontaneously breaks these symmetries. A $\mathcal{P}$ transformation along the flat directions (say, in Poincar\'{e} patch) of the $AdS_4\times CY_3$ solution will flip the fluxes as
\begin{equation} \int F_4\,\rightarrow -\int F_4,\quad \int H_3\rightarrow-\int H_3,\quad m\,\rightarrow\, -m.\label{e22}\end{equation}
This is a \emph{different} solution of the DGKT family, which preserves SUSY if the original one does. All of its properties are identical to those of the original DGKT solution. An internal symmetry of the bosonic, massless IIA equations of motion that agrees with \eq{e22} (without including any spacetime action) was discussed in \cite{Marchesano:2020qvg,Marchesano:2021ycx}. We now see it as a consequence of the exact $\mathcal{P}$ symmetry of massless IIA. Indeed, when a symmetry is spontaneously broken, it maps the vacuum to a different one with the same properties, which is what we are seeing here. If one is only concerned with the vacuum solution, the parity transformation acts trivially, and one can forget about it, yielding the symmetry of \cite{Marchesano:2020qvg,Marchesano:2021ycx}. However, if one extends the analysis to fluctuations around the solution (which will be sensitive to the parity symmetry), only the full parity transformation will be an actual symmetry of the lagrangian.

The $\mathcal{C}$ transformation preserves $H_3$, but flips both Roman's mass $m$ and the $F_4$ flux,
\begin{equation} \int F_4\,\rightarrow -\int F_4,\quad m\,\rightarrow\, -m.\label{e23}\end{equation}
This does not preserve the tadpole \eq{tad2}, so something else must be changing too.  Indeed, the spin lift of the orientifold involution $\iota_3 \mathcal{C}$ does not commute with $\mathcal{C}$; rather, it anticommutes when acting on fermions. To see this, we can picture $\iota_3$ as the composition of three reflections $\mathcal{P}$, which at the level of fermions, are implemented by the product of three $\Gamma$ matrices. Since $\mathcal{C}$ is also implemented by a $\Gamma$ matrix ($\Gamma_{11}$), it anticommutes with $\mathcal{C}$. As a result, acting with $\mathcal{C}$ also changes the action on fermions on orientifold involutions, introducing an additional factor of $(-1)^F$. This is the same as saying that $\mathcal{C}$ changes the choice of Pin$^-$ structure on the orientifold background, effectively replacing it by an $\overline{O6}$, which has the opposite RR charge. The tadpole is thus preserved and the result of this transformation is to flip the chirality of the whole solution.

We also note in passing that we could not find an exact symmetry that flipped $\int F_4$, and nothing else. This class of deformations, known as ``skew-whiffing'' (see e.g. \cite{Giri:2021eob}), are often symmetries of the 10d equations of motion, but they  break supersymmetry, albeit only when quantum effects are taken into account. Since we are dealing only with exact symmetries, that commute with the supercharges, we will not be able to land into a SUSY-breaking vacuum if we start with a SUSY-preserving one.

The symmetries discussed above are the only universal ones in a type IIA compactification with fluxes. There could be additional reflections of the $CY_3$, which could become symmetries at low energies if they commute with the isometries that define the $O6$ planes and also preserve all fluxes. These are non-generic, and must be analyzed in a case-by-case basis; furthermore, it is natural to expect that they will be broken in general by the ten-dimensional backreaction of the supergravity fields, which we discuss in Section \ref{sec:f33}. At any rate, even ignoring this parity-breaking backreaction, we will now show that there are no candidates in the benchmark DGKT model of Section \ref{sec:s2}, focusing on Pin$^+$ actions, as these are the only ones that are compatible with 3d $\mathcal{N}=1$ supersymmetry. 

To discuss the parity symmetries, we will first neglect fluxes, orientifold, and the action \eq{trolo} that turns the space into a Calabi-Yau, and just look at parity symmetries of IIA on $T^6$ that commute with the orientifold action. When acting on fermions, a Pin$^+$ action must be implemented by acting with either one or four gamma matrices. The reflection involving just one gamma matrix is the one described in the previous paragraphs, and is broken by fluxes. To discuss the actions involving four gamma matrices, it is useful to go to an M-theory picture, where all parity symmetries can be constructed out of reflections of various directions of an internal $T^7$. Of course, this picture is gone once Roman's mass is turned on; but because the massless and massive IIA fermion content is identical, the conclusions are still valid.

We will take coordinates 0,1,2,3 to describe the non-compact directions, 4,5,6,7,8,9 to describe $T^6$ ($z_1\sim x_4+ix_5$, etc), and finally, a 10th fictitious coordinate will describe the M-theory circle. In this picture, the O6 involution is generated by reflections in directions 5,7,9,10 where the extra reflection of the 10th coordinate accounts for the additional action of $(-1)^{F_L}\Omega$ described above. The action on fermions is via the product of gamma matrices
\begin{equation} \Gamma_{O6}=\Gamma_{5}\Gamma_{7}\Gamma_9\Gamma_{10}.\label{gammao6}\end{equation}
The reflection around say the $x_1$ direction is obtained as 
\begin{equation}\Gamma_{R}=\Gamma_1 \Gamma_{i}\Gamma_{j}\Gamma_{k},\end{equation}
where $\{i,j,k\}$ is a subset of $\{4,\ldots 10\}$ chosen such that $\Gamma_{R}$ commutes with \eq{gammao6}. For this to be the case, $\{i,j,k\}$ must contain an even number of elements in $\{5,7,9,10\}$. There are a few possible cases: \begin{itemize}
\item For $\{i,j,k\}$ where $i,j$ are in $\{5,7,9\}$ and $k=4,6,8$, the resulting parity symmetry is broken (at least) by the $F_4$ flux, which has legs in the 4567, 4589 and 6789 directions.
\item For $\{i,j,k\}=\{4,6,8\}$ (or permutations), the parity symmetry is simply 
\begin{equation} z_i\,\rightarrow -\bar{z}_i,\end{equation}
i.e. the same as \eq{ori9} except for the minus sign. This transformation preserves $m$, $H_3$, and $F_4$, but is broken by the orbifold \eq{trolo}, specifically because it does not commute with $z_i\,\rightarrow \alpha^2 z_i$. 
\item For $\{i,j,k\}=\{10,j,k\}$ where $j\in\{4,6,8\}$ and $k\in\{5,7,9\}$ (or permutations), the parity symmetry is preserved by the fluxes but again broken by the orbifold \eq{trolo}, this time because it does not commute with the first action. 
\end{itemize}

The summary is that the DGKT background spontaneously breaks all the universal, exact parity symmetries of massive IIA theory, and the benchmark DGKT scenario breaks all of the parity symmetries that would exist in the parent $T^6$. This also means that these symmetries should also be broken in the worldvolume theory of any brane probing this background\footnote{Often, the IR description of branes in string compactifications develops accidental global symmetries, which do not correspond to exact symmetries of the string background. See \cite{Kim:2019ths} for an example. However, the expectation would be that these symmetries should be broken as we go up the RG flow and start including irrelevant higher-derivative terms, etc.}. This lack of parity symmetry will have important consequences for the DGKT vacuum, as we will see. The worldvolume field theory will not be protected from receiving quantum corrections, which will lift the moduli space and render the AdS vacuum in tension with the WGC.

\section{Fate of the DGKT moduli space at quantum level\label{sec:modquantum}} 
To describe the moduli space of the DGKT theory, we must understand the dynamics of branes in this geometry in detail. We will do so in this Section, first studying the classical theory of a D4 brane away from a singularity, then moving on to the theory at singularities, and finally, understanding the quantum effects on these. 

\subsection{Worldvolume theory of the wrapping D4-brane} \label{sec:ang}

As described in \cite{DeWolfe:2005uu,Aharony:2008wz} and above, the DGKT construction is a parametric family of 4d $\mathcal{N}=1$ AdS$_4$ vacua, labelled by the value of 4-form flux on a particular 4-cycle of the internal $CY_3$,
\begin{equation} N=\int_{\omega_4} F_4.\end{equation}
By the completeness principle, one expects to have BPS domain walls that mediate flux jumps by one unit. The natural candidate are $D4$-branes wrapping the  2-cycle $\omega_2$ dual to $\omega_4$. The classical conditions for supersymmetry of D4 branes were analyzed in \cite{Aharony:2008wz}: this 2-cycle needs to be holomorphic. The worldvolume theory of these domain walls, describing their fluctuations at long wavelenghts, must be a 3d $\mathcal{N}=1$ quantum field theory, since this is the amount of supercharges preserved by a 4d $\mathcal{N}=1$ BPS domain wall. 

In this Section, we will construct and analyze in detail the worldvolume theory of a single D4 brane wrapping a 2-cycle dual to the $F_4$ flux in the benchmark DGKT model of Section \ref{sec:s2}. An analysis of the DGKT moduli space of this model via wrapped D4 branes was already carried out in \cite{Aharony:2008wz}; however, that reference did not explore the physics of D4 branes in singularities, which is where the most interesting effects happen.

Let us start with the case where the D4 brane is away from the orientifold locus, so that the worldvolume field theory can be simply obtained by dimensional reduction of the worldvolume theory of a $D4$ brane wrapping a 2-cycle (as in \cite{Aharony:2008wz}). In Subsection \ref{sec:map} we will describe how this field theory changes depending on the exact location of the brane.

The worldvolume action for a $Dp$-brane, with worldvolume coordinates $\xi^0,\xi^1,\ldots \xi^{p}$,  famously consists of two terms, the Dirac-Born-Infeld (DBI) and Chern-Simons pieces (see e.g. \cite{polchinski1998string}),
\begin{equation}S_{\text{brane}}= -T_{\text{brane}} \int_{\text{brane}} d^{p+1} \xi\, e^{-\phi}\sqrt{-\text{det}(h+\mathcal{F})} + iQ_{\text{brane}} \int_{\text{brane}} [C \wedge e^{\mathcal{F}}]_{p+1}.\label{jaja}\end{equation}
Using the conventions of \cite{polchinski1998string} (which differ from those of \cite{DeWolfe:2005uu,Aharony:2008wz} by a factor of $\sqrt{2}$, see footnote 1 in \cite{Aharony:2008wz} and footnote 3 in \cite{DeWolfe:2005uu}), we have
\begin{equation}\label{TQ} T_{\text{brane}}^2=Q_{\text{brane}}^2=\frac{\pi}{\kappa_{10}^2}(4\pi^2\alpha')^{3-p},\quad \kappa_{10}^2=\frac12(2\pi)^7(\alpha')^4.\end{equation}
The first term of \eq{jaja} (the DBI piece) contains kinetic terms for the $(9-p)$ scalars parametrizing the transverse directions to the brane, via the induced metric
\begin{equation}h_{ab}= g_{\mu\nu}\frac{\partial X^\mu}{\partial \xi^a}\frac{\partial X^\mu}{\partial \xi^b}\end{equation}
which is just the pull-back of the ambient 10d metric $g_{\mu\nu}$ (in string frame). Defining $\mathcal{F}_{ab}\equiv \pi \alpha'\, F_{ab}+ B_{ab}$, the DBI action is seen to also include kinetic terms for the worldvolume $U(1)$ gauge field $A_a$, with fieldstrength $F_{ab}$.  The DBI term also includes a coupling to the pullback of the 10d NS NS field $B_{\mu\mu}$, via its pullback to the worldsheet $B_{ab}$.

The second term (the Chern-Simons piece) encodes the D-brane coupling to RR fields. Expanding it for the particular case of a $D4$ brane, we have
\begin{equation} \label{CSexp} \int_{\text{brane}} [C \wedge e^{\mathcal{F}}]_{p+1}=\int  C_5 + \int C_3 \wedge \mathcal{F} + \int C_1 \wedge \frac{\mathcal{F}^2}{2} + \frac{m}{6} \int \mathcal{F} \wedge\mathcal{F} \wedge A,\end{equation}
where the last term (the coupling to Romans' mass) can be formally understood as coming from integration by parts of a formal expression proportional to $C_{-1} \wedge \mathcal{F}^3$. 

The action \eq{jaja} is valid to all orders in $\alpha'$ expansion (for constant $F$) and small $g_s$.   We will now particularize the analysis to the $D4$ brane we are interested in. We will consider AdS$_4$ in the Poincar\'{e} patch, with coordinates
\begin{equation}ds^2=\frac{R^2}{\ell_{\rm AdS}^2}\left(-dX_0^2+dX_1^2+dX_2^2\right)+ \frac{\ell_{\rm AdS}^2}{R^2}dR^2\,\end{equation}
 and a $D4$-brane sitting at a particular value $R=R_0$ of the radial coordinate, and wrapping the internal cycle $\omega_2$. This configuration is classically supersymmetric, as studied in \cite{Aharony:2008wz}. We will take the gauge
 \begin{equation} \xi^0=X_0,\quad \xi^i=X^i,\quad i=1,2, \quad \xi^i=z^i\quad i=3,4,\end{equation}
 where $z^3,z^4$ are two coordinates parametrizing the cycle $\omega_2$. The brane is localized in the representative of the cycle $\omega_2$ that minimizes its action; the only other function needed to specify the embedding is $R(\xi^a)$, specifying fluctuations of its position in the  normal AdS direction.
 
  In the DGKT scenario, as $N$ is taken to be very large, the cycle $\omega_2$ has a very large volume in string units \cite{DeWolfe:2005uu}. Therefore, as a first approximation, we will drop all $\alpha'$ corrections in the DBI term of the action, as these would yield corrections suppressed by the string scale. Furthermore, since we are only interested in IR dynamics at wavelengths much longer than the Kaluza-Klein scale of the internal manifold (and in particular, the characteristic size of $\omega_2$, which is comparable to it), we will perform a dimensional reduction on the cycle $\omega_2$. The five-dimensional bosonic worldvolume fields reduce as follows:
 
\begin{itemize}
\item From the five-dimensional $U(1)$ gauge field, we obtain a 3d $U(1)$ gauge field $A$, and Wilson line scalars $a_i$ valued in $H^1 (\omega_2,\mathbb{R})$ (ignoring large gauge transformations). For a torus we get two scalars, $a_1,a_2$ . 
\item The  five-dimensional scalars parametrizing the normal coordinates to the $D4$ brane in ten dimensions split as the scalar $R$, as well as four scalars $u^i$ describing the precise location of the $D4$ in the internal space. In toroidal DGKT examples, these are locally four real scalars with a flat metric \cite{Aharony:2008wz}, and they all survive at low energies.\end{itemize}

We also need to perform the dimensional reduction on the 10d background fields restricted to the $D4$ worldvolume, since these affect the couplings of the theory. The $B_2$ field is odd under the orientifold involution present in the DGKT model, while the $D4$ brane wraps even cycles. As a result, the restriction of $B_2$ decomposes in terms of orientifold-odd KK modes, which have zero overlap with the zero modes of the bosonic fields we discussed above. Therefore, in practice, we can take $B_2=0$ in \eq{jaja} to obtain the 3d effective action. Similarly, in simple versions of the DGKT scenario, both $C_1$ and $C_3$ are not turned off. Including backreaction can result in cohomologically trivial nonzero $C_1$,$C_3$ profiles turned on \cite{Marchesano:2020qvg}, but when integrated on $\omega_2$, these must always vanish. Therefore, again at the level of the low-energy EFT of the $D4$ brane, we might as well assume $C_1=C_3=0$. 

With these observations, we can now easily write down the dimensional reduction of \eq{jaja}. We have the induced 5d metric (after integrating over the compact coordinates)
\begin{equation} h_{D4}= \frac{R^2}{\ell_{\rm AdS}^2}(-d\xi_0^2+d\xi_1^2+d\xi_2^2)+ \frac{\ell_{\rm AdS}^2}{R^2}(\vec{\nabla} R \cdot d\vec{\xi}\, )^2 + \sum_{i,j} g^{CY_3}_{ij} (\vec{\nabla} u^i \cdot d\vec{\xi}\, )(\vec{\nabla} u^j \cdot d\vec{\xi}\, ), \end{equation}
where $g^{CY_3}_{ij}$ is the moduli space metric obtained from integrating the Calabi-Yau metric for the normal bundle coordinates and integrating over $\omega_2$, as well as the gauge field 
\begin{equation} F^{5d}= F + \sum_{i\in H^1(\omega_2,\mathbb{R})} d a_i\wedge \chi^i,\end{equation}
where $\{\chi^i\}$ are a basis of de Rham representatives of all cohomology classes in $H^1 (\omega_2,\mathbb{R})$. For the $T^2$ case that we are focusing on, there are two such modes. We can now evaluate the determinant in the DBI action, assuming all perturbations are small\footnote{Using the formula  $\text{det}(\mathbf{I}+\mathbf{A})\approx 1+ \text{tr}(\mathbf{A})+ \frac{\text{tr}(\mathbf{A})^2-\text{tr}(\mathbf{A^2})}{2}+\ldots$}, obtaining
\begin{align}-&\text{det}(h_{D4}+ 2\pi \alpha' F^{5d})\approx\nonumber\\&\frac{R^6}{\ell_{\rm AdS}^6}\left[1+ \frac{\ell_{\rm AdS}^4}{R^4} \vert \nabla R\vert^2+\frac{\ell^2_{\text{AdS}}}{R^2}\sum_{i,j} g_{ij}^{CY_3} \nabla u^i\cdot \nabla u^j + (2\pi \alpha')^2\frac{\ell_{\rm AdS}^4}{2R^4}\left(\vert F\vert^2+ \vert \nabla a_i\vert^2\right) \right]\text{det}(g^{T^2})\end{align}
plus higher-derivative terms. This translates in a DBI contribution to the action of
\begin{align}\frac{S_{\text{DBI}}}{\text{Vol}(\omega_2)}= &T_{\text{brane}} \int d^3\xi \,g_s^{-1}\left( -\frac{R^3}{\ell_{\rm AdS}^3}-\frac{\ell_{\rm AdS}}{2R} \vert \nabla R\vert^2-\frac{R}{2\ell_{\rm AdS}} \sum_{i,j} g_{ij}^{CY_3} \nabla u^i\cdot \nabla u^j -\right.\nonumber\\&\left.-\frac{(2\pi \alpha')^2\ell_{\rm AdS}}{4R}\vert F\vert^2-\frac{(2\pi \alpha')^2R}{4\ell_{\rm AdS}} \vert \nabla a_i\vert^2\right).\label{e344}\end{align}

Turning now to the Chern-Simons terms, one just needs to substitute the profile for $C_5$ in \eqref{CSexp}, which comes from the pullback of the background $F_6$ threading spacetime, 
\begin{equation} 
F_6=\star F_4= \frac{R^2}{\ell_{\rm AdS}^2}f_1\frac{\gamma_1}{\gamma_2\gamma_3}d^3\xi dR\, du_1\,dv_1
\end{equation} 
where we have used that $F_4=f_i\hat\omega_i$. Here, for simplicity, we have particularized to the concrete CY in \eqref{cy3}, using  $ds_{T^6}^2=\sum_{i=1}^3\gamma_i(dx_i^2+dy_i^2)$ as the metric for the covering toroidal space.
The result is simply
\begin{equation} \frac{S_{\text{CS}}}{\text{Vol}(\omega_2)}=Q_{\text{brane}}\left(\int d^3\xi \frac{R^3}{3\ell_{\rm AdS}^4} f_1\frac{1}{\gamma_2\gamma_3} + \frac{m}{6}(2\pi \alpha')^3\int d a_1 \wedge d a_2 \wedge A\right)\ .\label{simons}\end{equation} 
The first term is equal and opposite to the tension contribution in \eq{e344}, by virtue of the four-dimensional Einstein equations, which stabilize the moduli as follows
\begin{equation}
e^{-\phi}\sim \frac{1}{|h_3|}(m_0f_1f_2f_3)^{1/4}\ ;\ \gamma_i\sim \frac{1}{|f_i|}\ ; \ \ell_{\rm AdS}\sim \frac{h_3^2}{\gamma_1\gamma_2\gamma_3}\sqrt{\frac{f_1f_2f_3}{m_0}}
\end{equation}
where more details can be found in  \cite{DeWolfe:2005uu,Aharony:2008wz}.
The resulting cancellation is the no-force condition which is the hallmark of BPS branes. It was also checked in \cite{Aharony:2008wz}. This cancellation also implies that $R$ is a modulus at the classical level, but we do not expect this to be true after quantum corrections are taken into account.  Let us analyse the rest of the action. Upon adding up the Chern-Simons and DBI terms, we get the final following effective action
\begin{align}
S_{\text{brane}}&=  \int d^3\xi \,\left(T_{\text{brane}}\frac{\text{Vol}(\omega_2)}{g_s}\left(- \frac{\ell_{\rm AdS}}{2R} \vert \nabla R\vert^2-\frac{R}{2\ell_{\rm AdS}} \sum_{i,j} g^{CY_3} \nabla u^i\cdot \nabla u^j -\right.\right.\nonumber\\&\left.\left.-\frac{(2\pi \alpha')^2R}{4\ell_{\rm AdS}}\vert \nabla a_i\vert^2\right)-\,\frac{1}{2g^2_{YM}}\vert F\vert^2\,+\,\frac{m}{6}Q_{\text{brane}}(2\pi\alpha')^3\text{Vol}(\omega_2)\int d a_1 \wedge d a_2 \wedge A\right),\label{effaction}\end{align}
where the gauge coupling is given by
\beq
g^2_{YM}=\frac{2R}{\ell_{\rm AdS}} (T_{\rm{brane}}\,g_s^{-1}\text{Vol}(\omega_2)(2\pi\alpha')^2)^{-1}=\frac{R}{\ell_{\text AdS}}\frac{8\pi^2g_s\alpha'^{1/2}}{\text{Vol}(\omega_2)}\ .
\eeq

All we have are, thus,  $R$-dependent kinetic terms for all scalar fields and the $U(1)$ gauge field. Since there is no potential for $R$, the vev $R_0$ is a modulus at the classical level and parametrizes a family of supersymmetric ground states of the worldvolume QFT. This implies that the brane feels no force (at the classical level) and can be placed at any value of $R_0$ from the Poincare horizon. This was the result of \cite{Aharony:2008wz}. However, as discussed in Section \ref{sec:parDGKT}, the DGKT vacuum has no parity symmetries and, therefore, we expect this moduli space to be lifted by quantum corrections. In other words, we expect this scalar $R$ to get a potential. We will show this soon, but first, we must describe the theory the D4 branes close to the orbifold and orientifold singularities of the DGKT geometry. 

\subsection{A global picture of the D4 brane moduli space and IR EFT}\label{sec:map}
Before we can see in detail how quantum effects lift the moduli space of the D4 brane, we first need to describe all branches of the classical moduli space of the brane in some detail. Consider a D4 brane wrapped on the holomorphic 2-cycle spanned by one of the $T^2$ factors in the covering $T^6$ of the Calabi-Yau, say the one parametrized by $z_1$.  On top of the Wilson line moduli, there are two complex scalars  parametrizing the position of the brane on the other two $T^2$ factors $z_2,z_3$ of the covering $T^6$, as described in the previous Subsection.

 For generic values of $z_2,z_3$ the orbifold and orientifold projections identify this $T^2$ with distinct orbifold/orientifold images, and the resulting brane still wraps a $T^2$ on the $CY_3$.  The dimensional reduction discussed in the previous Section corresponds to this first case, where the brane is wrapping a generic $T^2$ cycle. The dimensional reduction goes exactly as it would in flat space, and since the tension and gauge field contributions cancel, the net result in the deep IR is that one obtains, classically, a 3d $\mathcal{N}=8$ theory, containing the same zero modes and gauge fields that one would obtain from dimensional reduction of a $D4$ on $T^2$ (namely, eq.\eqref{effaction}). The additional effects coming from e.g. Roman's mass, such as the topological coupling proportional to \eq{CSexp} and its supersymmetric counterparts, are suppressed by the string scale and become irrelevant in the deep IR. This is just encoding the usual fact that a brane in a compact space is only probing its local geometry, and in this case, the local geometry is flat.

If $(z_2,z_3)$ are tuned to different values, however the D4 brane worldvolume may intersect the orientifold or the orbifold locus, and this affects the low-energy EFT even in the deep IR. The relative geometry between D4 and O6$^{-}$ plane\footnote{The DGKT construction requires the use of $O6^-$ planes \cite{DeWolfe:2005uu}, since $O6^+$ would require a sign flip in the quantity $mH_3$ due to the tadpole \eq{tad2}, which is incompatible with supersymetry. One easy way to see this is to use the symmetry $\mathcal{C}$ of M-theory described in Section \ref{sec:parDGKT} on the standard DGKT solution, which flips $m H_3$ and sends an O6$^-$ to an $\overline{\text{O6}^-}$. $mH_3$ now has the right sign that would be required to cancel an O6$^+$ tadpole, but the supersymmetry preserved is opposite.}  is
\begin{equation}
\begin{array}{c|cccccccccc}
\text{Brane}&T&X_1&X_2&R&\text{Re}(z_1)&\text{Im}(z_1)&\text{Re}(z_2)&\text{Im}(z_2)&\text{Re}(z_3)&\text{Im}(z_3)\\\hline
D4& \textrm{--}&\textrm{--}&\textrm{--}&\times&\textrm{--}&\textrm{--}&\times&\times&\times&\times\\
O6& \textrm{--}&\textrm{--}&\textrm{--}&\textrm{--}&\textrm{--}&\times&\textrm{--}&\times&\textrm{--}&\times\\
\end{array}
\label{geom9}\end{equation}
We see that the D4 is transversal to the O6 in the $z_1$ direction.
Hence, for instance, when both $z_2$ and $z_3$ are purely real (but at least one of Re$(z_2)$, Re$(z_3)$ is nonzero), the D4 brane worldvolume intersects the O6 locus at Im$(z_1)=0$. This means that the D4 brane coincides with its image in the covering space, and also that the orientifold action projects out some of the degrees of freedom (and reduces the degree of supersymmetry) of the D4 brane stack.
More concretely, an O6$^-$ plane imposes an orthogonal projection on D-branes parallel to it, but on branes that intersect it transversally, the orientifold projection is instead symplectic\footnote{Although the analysis is a bit telegraphic, reference \cite{Aharony:2008wz} seems to take an orthogonal orientifold projection instead. This has important consequences for the IR dynamics of the D4 brane, which is why we belabor the point.}\cite{Gimon:1996rq}. Perhaps the quickest way to see this for our stack is to T-dualize along the $\text{Im}(z_1),\text{Im}(z_2),\text{Im}(z_3)$ coordinates to a D5-O9 system, in which the orientifold projection is famously symplectic \cite{Witten:1995gx,Gimon:1996rq}; see \cite{Pradisi:1988xd} for an earlier discussion of these models. A general discussion of branes intersecting orientifolds transversely can be found in \cite{uranga-lectureTASI}, and the additional sign on the orientifold action responsible for this change is discussed in \cite{Gimon:1996rq} from a worldsheet perspective.

Finally, if $z_2$ and $z_3$ both vanish (or are fixed at the location of any of the other orbifold fixed points), the D4 brane will sit at an orbifold fixed point. All nine orbifold fixed points have an orientifold going through them, so the D4 at an orbifold fixed point also necessarily intersects the O6 locus. The geometry of the orbifold fixed locus is $T^2/\mathbb{Z}_3$, and the orbifold action will also change the EFT.

In the following, we work out the $D4$-brane field theory in detail near one of the singularities. We will first consider the case where the D4 brane is located on top of both the orientifold and orbifold fixed locus; after we work out the corresponding low-energy effective theory, the other cases will come out from higgsing by various fields.

\subsubsection*{Orbifold action}

For concreteness, we will take the brane to be located at $z_2=z_3=0$, wrapping the $z_1$ torus. As a first step, we will study the orbifold action on the low energy EFT, and add the orientifold projection later on. The bosonic worldvolume fields are $\mathbf{R}, \mathbf{A}_\mu$, $\mathbf{\Phi}_2$, and $\mathbf{\Phi}_3$, where $\mathbf{\Phi}_2$ and $\mathbf{\Phi}_3$ parametrize deviations in the $z_2,z_3$ planes respectively, and $\mathbf{R}$ parametrizes the position of the brane in the AdS radial direction (being extended in the transverse coordinates). The boldface indicates that all these fields carry Chan-Paton labels and are in the adjoint representation of $U(N)$, where $N$ is the number of branes we put in the covering space -- before taking into account orbifold and orientifold corrections. As will become clear later, the minimal $N$ in order to have mobile branes is $N=6$, but we will keep $N$ generic for the time being.

With these definitions,the orbifold projection imposes the identifications (for mobile branes)
\begin{align} \mathbf{\Phi}_i(\vec{x},z_1)&=\alpha^2 \mathbf{P}_3\mathbf{\Phi}_i(\vec{x},\alpha^2 z_1)\, \mathbf{P}_3^{-1},\quad \mathbf{R}(\vec{x},z_i)= \mathbf{P}_3\mathbf{R}(\vec{x},\alpha^2 z_1)\, \mathbf{P}_3^{-1},\nonumber\\ \mathbf{A}_\mu(\vec{x},z_1)&= R_\mu^\nu\, \mathbf{P}_3\mathbf{A}_\nu(\vec{x},\alpha^2 z_1)\, \mathbf{P}_3^{-1},\label{orbip1}\end{align}
where $\mathbf{P}_3$ is a $U(N)$ gauge transformation that acts on the fundamental Chan-Paton factors via a permutation matrix\footnote{The gauge transformation $\mathbf{P}_3$ is taken in e.g. the fundamental representation; hence, its action in adjoint fields is by left and right multiplication.} \cite{Johnson:2000ch},
\begin{equation}\mathbf{P}_3=\left(\begin{array}{ccc}
 0 & \mathbf{I}_{N/3\times N/3} & 0 \\
 0 & 0 & \mathbf{I}_{N/3\times N/3} \\
 \mathbf{I}_{N/3\times N/3} & 0 & 0 \\
\end{array}
\right).\end{equation}
For this element to exist, we must take $N$ to be a multiple of 3. Any such element is essentially unique, since it is conjugate to the transformation that groups the Cartans in three blocks of $N/3$ elements, and permutes them. The role of $\mathbf{P}_3$ can be understood clearly in the Coulomb branch: There are three separate stacks of $D4$-branes, and $\mathbf{P}_3$ simply permutes one brane to the next as we implement the orbifold action. The factor of $\alpha^2$ in the action of the $\mathbf{\Phi}_i$ is also geometrically clear -- it implements a 120$^\circ$ rotation in the $z_2,z_3$ planes, and the rotation matrix $ R_\mu^\nu$  implements the rotation in the $z_1$ plane corresponding to the orbifold action, as well. 

The role of \eq{orbip1} is to project out a subsector of the $T^2$ KK modes. Without this projection, we would have obtained a 3d $\mathcal{N}=8$ theory -- simply the dimensional reduction of 5d $\mathcal{N}=2$ SYM theory on a torus. Let us now compute the bosonic zero modes of the compactification. The adjoint of $U(N)$ equals a product of a fundamental and an anti-fundamental. In the fundamental representation, the action of $\mathbf{P}_3$ amounts to simply permuting coordinates, and commutes with a $U(N/3)^3$ subgroup of $U(N)$. Upon the decomposition $U(N)\,\rightarrow\, U(N/3)^3\times\mathbb{Z}_3$, we have
\begin{align} \mathbf{N}&\rightarrow\, \left(\mathbf{N/3},\mathbf{1},\mathbf{1}\right)_{0}\oplus  \left(\mathbf{1},\mathbf{N/3},\mathbf{1}\right)_{1}\oplus  \left(\mathbf{1},\mathbf{1},\mathbf{N/3}\right)_{2},\nonumber\\ \bar{\mathbf{N}}&\rightarrow\, \left(\bar{\mathbf{N}}\mathbf{/3},\mathbf{1},\mathbf{1}\right)_{0}\oplus  \left(\mathbf{1},\bar{\mathbf{N}}\mathbf{/3},\mathbf{1}\right)_{2}\oplus  \left(\mathbf{1},\mathbf{1},\bar{\mathbf{N}}\mathbf{/3}\right)_{1},\end{align}
where the subscript denotes the $\mathbb{Z}_3$ charge. This means that, under the same decomposition,
\begin{align} \text{Adj}_{U(N)}=\mathbf{N}\otimes\bar{\mathbf{N}}&\rightarrow\, \left(\text{Adj}_{U(N/3)},\mathbf{1},\mathbf{1}\right)_0\oplus \left(\mathbf{1},\text{Adj}_{U(N/3)},\mathbf{1}\right)_0\oplus \left(\mathbf{1},\mathbf{1},\text{Adj}_{U(N/3)}\right)_0\nonumber\\&\oplus \left(\mathbf{N/3},\bar{\mathbf{N}}\mathbf{/3},\mathbf{1}\right)_2\oplus\left(\mathbf{1},\mathbf{N/3},\bar{\mathbf{N}}\mathbf{/3}\right)_2\oplus\left(\bar{\mathbf{N}}\mathbf{/3},\mathbf{1},\mathbf{N/3}\right)_2\nonumber\\\nonumber\\&\oplus \left(\mathbf{1},\bar{\mathbf{N}}\mathbf{/3},\mathbf{N/3}\right)_1\oplus \left({\mathbf{N}}\mathbf{/3},\mathbf{1},\bar{\mathbf{N}}\mathbf{/3}\right)_1\oplus
\left(\bar{\mathbf{N}}\mathbf{/3},\mathbf{N/3},\mathbf{1}\right)_1,\label{lelian}\end{align}
and since the zero modes pick the singlet part under the permutation $\mathbb{Z}_3$, we get fields in the adjoint of $U(N/3)^3$. 
Another way to see this is to notice that $\mathbf{P}_3$ is conjugate to a Cartan element, and every other direction in the same Cartan algebra will commute with it; since there are $N$ of those, the rank has to remain the same. Therefore, at the level of zero modes, the effect of the orbifold is to project out the gauge group as
\begin{equation}U(N)\,\rightarrow\, U\left(\frac{N}{3}\right)^3.\end{equation}
Since the field $\mathbf{R}$ transforms under $\mathbf{P}_3$ in the same way as the $\mu=0,1,2$ components of the gauge field, it is projected out in the same way, producing three zero modes valued in the adjoint of each $U(N/3)$ factor.

The fields  $\mathbf{A}_{4,5}$ (the components of the gauge field along the $T^2$ directions) transform into each other due to the 120$^\circ$ rotation in \eq{orbip1}, acting on the 2-dimensional real representation. These can be complexified into a single complex field transforming similarly to the $\mathbf{\Phi}_i$ in \eq{orbip1}, so we can analyze all of them simultaneously. The extra phase factor in \eq{orbip1} forces us to choose a non-zero eigenvalue, picking either the second or third lines of \eq{lelian} (it does not matter which, since they are all related by charge conjugation). The resulting zero modes
\begin{equation} ((\mathbf{A}^{(0)})_\mu,\mathbf{R}^{(0)}),\quad (\mathbf{\Phi_2}^{(0)}),\, (\mathbf{\Phi_3}^{(0)}),\quad\text{and}\quad (\mathbf{A}_{4,5}^{(0)}) \end{equation}
comprise the bosonic content of a 3d $\mathcal{N}=2$ vector multiplet in the adjoint of $U(N/3)^3$, and three 3d $\mathcal{N}=2$ chiral multiplets in bifundamental representations \cite{Aharony:1997bx,Johnson:2000ch}. Thus, the orbifold breaks the supersymmetry to 3d $\mathcal{N}=2$. This matches the analysis of the supercharges: The 5d $\mathcal{N}=2$ supercharge transforms in the $(\mathbf{4},\mathbf{4})$ of Spin(5,1)$\times$ Sp(2), subject to a reality condition. The local Calabi-Yau geometry breaks the above as
\begin{equation} \text{Spin}(5,1)\times\text{Sp}(2)_R\,\rightarrow\, \text{Spin}(2,1)\times U(1)\times U(1)_R\times SU(2)_R,\end{equation}
where the $T^2/\mathbb{Z}_3$ geometry implies a nontrivial bundle for both $U(1)$ and $U(1)_R$. Under this decomposition the supercharge decomposes as (the subindex denotes charge under $U(1)+U(1)_R$)
\begin{equation} (\mathbf{4},\mathbf{4})\,\rightarrow\, (\mathbf{2},\mathbf{2})_{\frac12}+(\mathbf{2},\mathbf{2})_{-\frac12}+(\mathbf{2},\mathbf{1})_{1}+(\mathbf{2},\mathbf{1})_{-1}+2\,(\mathbf{2},\mathbf{1})_{0},\end{equation}
Upon imposing the reality condition, the only surviving zero mode is a single copy of a complex $(\mathbf{2},\mathbf{1})_{0}$, which amounts to four supercharges\footnote{There is an alternative breaking pattern where the $(\mathbf{4},\mathbf{4})$ goes, after imposing the reality conditon, to the combination $(\mathbf{2},\mathbf{2})_0+(\mathbf{2},\mathbf{2})_1$, so there are eight surviving supercharges in the $(\mathbf{2},\mathbf{2})$ and the compactification has 3d $\mathcal{N}=4$ supersymmetry \cite{Yonekura:2013mya}. This would describe a case with accidental SUSY enhancement; a breaking to four supercgarges (3d $\mathcal{N}=2$) is what one would generically expect from the Calabi-Yau geometry.}.  We can also reproduce directly the counting of hypermultiplets, which come from normal deformations of the $D4$ brane which transform in the $\mathbf{5}$ of the $Sp(2)$ 5d R-symmetry group. Under the same decomposition, the $\mathbf{5}$ of Sp(2) decomposes as
\begin{equation}\mathbf{5}\,\rightarrow\, \mathbf{2}_{\frac12}+\mathbf{2}_{-\frac12}+\mathbf{1}_0,\end{equation}
which again agrees with the fact that there is a single massless scalar coming from the normal deformation in the $\mathbf{R}$ direction, which does not require a gauge bundle, and two normal $\mathbf{\Phi}_i$ deformations coming from either $ \mathbf{2}_{\frac12}$ or $\mathbf{2}_{-\frac12}$ -- whichever of these matches the eigenvalue under the gauge transformation $\mathbf{P}_3$.

On top of the spectrum, we also obtain a scalar potential, coming from reduction from the 5d Yang-Mills kinetic term,
\begin{equation} V\sim \text{Tr}\left( \sum_i\,  [\mathbf{R}^{(0)}, \mathbf{\Phi}_i]^2+[\mathbf{R}^{(0)}, \mathbf{A}^{(0)}]^2+ [\mathbf{\Phi}^{(0)}_2, \mathbf{\Phi}^{(0)}_3]^2+  [\mathbf{\Phi}^{(0)}_i, \mathbf{A}^{(0)}]^2\right),\label{scapot}\end{equation} 
which comes from a cubic 3d $\mathcal{N}=2$ superpotential (see e.g. \cite{Uranga:2000ck}).
The scalar potential \eq{scapot} has two branches of vacua. On the Coulomb branch, the scalar  $\mathbf{R}^{(0)}$ gets a vev, while $\mathbf{\Phi}^{(0)}_i$ and $\mathbf{A}^{(0)}$ have vanishing vevs and are made very massive by the $\mathbf{R}^{(0)}$ vev. Each of the $U(N/3)$ factors can be broken independently, leading, on a generic point in the Coulomb branch, to $N$ decoupled  3d $\mathcal{N}=2$ sectors, each with a $U(1)$ gauge group, and no chiral multiplets. The physical interpretation is that the $N/3$ mobile D4 branes can fractionalize on the singularity, and each fractional brane has a charge 1/3 of a mobile brane.  Each of the fractional branes is then ``trapped'' in the singularity, with the position moduli being frozen and projected out by the combination of orbifold and orientifold actions. There is a nice geometrical perspective for the existence of these fractional branes. Topologically, $T^2/\mathbb{Z}_3$ is not a torus but a $\mathbb{P}^1$, and in particular it is rigid. Therefore, wrapping a $D4$ on it should not have any normal deformations \cite{Yonekura:2013mya}, which agrees with the counting for fractional branes.  On  the other hand, this $\mathbb{P}^1$ is triple-covered by a movable $T^2$, so wrapping the cycle three times should include normal deformations. 
On the so-called Higgs branch, the  $\mathbf{R}^{(0)}$ admits at most an identical vev in each of the $U(N/3)$ factors, and the $\mathbf{\Phi}^{(0)}_i$ and $\mathbf{A}^{(0)}$ can get arbitrary, commuting vevs. A non-zero vev signals a stack of branes away from the singularity and it breaks the gauge group to the diagonal $U(N/3)$, corresponding to $N/3$ branes moving away from the singularity. Each bifundamental gives rise to an adjoint. At low energies, the system on the Higgs branch flows to the 3d $\mathcal{N}=8$ theory of D4 branes wrapping a $T^2$ that describes the $D4$ branes on a generic point of the Calabi-Yau, away from the orbifold singularity.

\subsubsection*{Orientifold action}

As a next step, we impose the orientifold projection. The D4 branes intersect the orientifold (located at $\text{Im}(z_1)=0$) transversally, and, as described above, in the double-cover the system is described by a stack of branes (as opposed to a brane-antibrane pair), with a symplectic action on Chan-Paton factors \cite{Gimon:1996rq}. On the D4 fields (represented as $U(N)$-valued matrices) it acts as
\begin{align} &\mathbf{A}_\mu(\vec{x},z_1)\rightarrow\,   \,- \mathbf{\tilde{R}}\mathbf{M}\, \mathbf{A}^T_\nu  (\vec{x},\bar{z}_1)\, \mathbf{M}^{-1},\quad \mathbf{\Phi}_i(\vec{x},z_1)\,\rightarrow\,\mathbf{M}\, \Phi^\dagger_i(\vec{x},\bar{z}_1)\, \mathbf{M}^{-1},\nonumber\\ &\mathbf{R}(\vec{x},z_1) \rightarrow \mathbf{M}\,\mathbf{R}^T(\vec{x},\bar{z}_1)\, \mathbf{M}^{-1}\label{ori}\end{align}
where $ \mathbf{\tilde{R}}_\mu^\nu=\text{diag}(1,1,1,1,1,-1)$ is simply the action on the vector representation corresponding to the reflection $z_1\,\rightarrow\,\bar{z}_1$, and $\mathbf{M}$ is the Chan-Paton matrix (i.e. in the fundamental of $U(N)$) 
\begin{equation} \mathbf{M}=\left(\begin{array}{cc}0&-i\,\mathbf{I}_{N/2\times N/2}\\i\, \mathbf{I}_{N/2\times N/2}&0\end{array}\right),\quad \mathbf{M}^2=\mathbf{I}_{N\times N}.\label{odin}\end{equation}
The  $-1$ in front of the transformation law of the gauge field can be understood from open string perturbation theory \cite{Gimon:1996rq} (from the fact that the orientifold reverses the worldvolume orientation of the string, but does not flip the endpoints for strings attached to the D4 brane), or alternatively from the requirement that the D-brane Chern-Simons terms should be invariant under the orientifold action. For the element in \eq{odin} to be well-defined, $N$ must be even -- taken together with the orbifold restriction, we learn that $N$ must be a multiple of 6 if we are to have mobile branes. The additional action of complex conjugation on the $\mathbf{\Phi}_i$ is due to the fact that the orientifold flips two of the four normal coordinates to the brane (the imaginary parts of $z_2$ and $z_3$).

At low energies, the orientifold projects out some of the $T^2$ KK modes, just like the orbifold did. Ignoring the orbifold projection for a moment, the $D4/O6^-$ intersection by itself preserves 8 supercharges, since it is T-dual to a D5 in type I string theory as discussed earlier in this Subsection. We therefore obtain a 3d $\mathcal{N}=4$ theory at low energies, containing a $Sp(N/2)$ gauge group and a single hypermultiplet in the reducible two-index antisymmetric representation of  $Sp(N/2)$ \cite{ibanez2014string}. It is reducible because a two-index antisymmetric tensor $\Phi_{ab}$ can be taken out a trace with the antisymmetric tensor preserved by the symplectic structure to obtain the scalar $\epsilon^{ab}\Phi_{ab}$. In terms of the 5d gauge fields, the 3d vector comes from the $\mu=0,1,2$ components of the 5d gauge field. The scalar $\mathbf{R}$, together with the component of $A_\mu$ along the $\text{Im}(z_1)$ and the real parts of $\Phi_i$ comprise the four scalars in the antisymmetric tensor representation, while the component of $A_\mu$ along $\text{Re}(z_1)$ and the imaginary parts of the $\Phi_i$ provide the adjoint scalars. We remind the reader that the adjoint representation is the same as the two-index symmetric in a symplectic group.

\subsubsection*{Putting everything together}
We can now combine the orbifold and orientifold actions, to find out the low-energy effective field theory when the stack of branes is at $z_2=z_3=0$. The action of the orbifold and orientifold commute, and combining both actions one obtains a 3d $\mathcal{N}=1$ with gauge group $Sp(N/6)^3$ consisting of:\begin{itemize}
\item One 3d $\mathcal{N}=1$ vector multiplet in the adjoint of $Sp(N/6)^3$.
\item One 3d $\mathcal{N}=1$ complex chiral superfield in the reducible, antisymmetric tensor representation of $Sp(N/6)^3$,
\begin{equation} (\mathbf{Asym},\mathbf{1},\mathbf{1})\oplus (\mathbf{1},\mathbf{Asym},\mathbf{1})\oplus (\mathbf{1},\mathbf{1},\mathbf{Asym})\end{equation}
corresponding to the scalar $\mathbf{R}^{(0)}$.
\item Three 3d $\mathcal{N}=1$ real chiral superfield in the real, sum of bifundamentals representation of $Sp(N/6)^3$,
\begin{equation} \left(\mathbf{N/3},\bar{\mathbf{N}}\mathbf{/3},\mathbf{1}\right)\oplus\left(\mathbf{1},\mathbf{N/3},\bar{\mathbf{N}}\mathbf{/3}\right)\oplus \left(\bar{\mathbf{N}}\mathbf{/3},\mathbf{1},\mathbf{N/3}\right)\label{ewwew2}\end{equation}
coming from the scalars $\mathbf{\Phi}_2^{(0)},\mathbf{\Phi}_3^{(0)},\mathbf{A}_{4,5}^{(0)}$ of the orbifold theory.
\end{itemize}
These fields are further supplemented with a real superpotential, which is the truncation of the orbifold theory one to the surviving multiplets. The reality condition in \eq{ewwew2}, which comes from the orientifold action on the orbifold bifundamentals, is possible because the fundamental representation of $Sp(N)$ is pseudorreal, so an $Sp(N)\times Sp(N)$ bifundamental admits  a reality condition (this would not work for unitary groups, where bifundamental representations are complex).

\begin{figure}[htb!]
\begin{center}
\includegraphics[scale = 0.5]{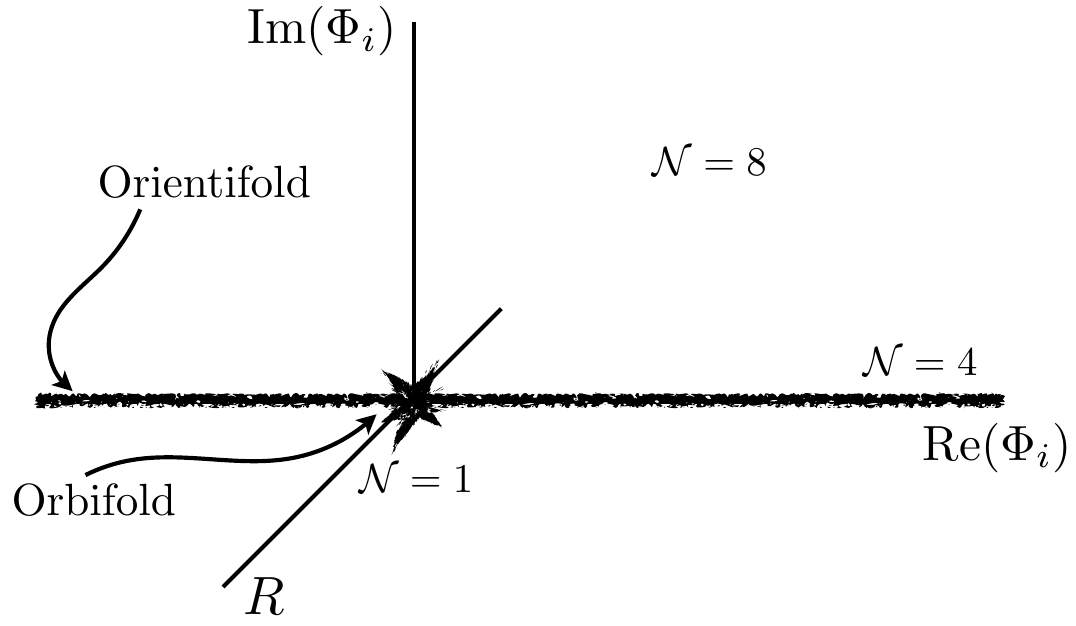}
\caption{Classical moduli space of D4 branes close to an orbifold-orientifold singularity in the original DGKT model of \cite{DeWolfe:2005uu}. The relevant moduli are the coordinate $R$ (radial AdS coordinate), and the complex scalars $\Phi_i$ (there are three of those, corresponding to two normal directions in the Calabi-Yau $\Phi_2,\Phi_3$ and one complexified Wilson line $\Phi_1\equiv A_4+i A_5$; we only depict one). The orbifold singularity sits at $\Phi_i=0$, and the orientifold corresponds to the $\text{Re}(\Phi_i)=0$ line. When $N/6$ mobile D4's are on top of the orbifold-orientifold intersection, the gauge theory is 3d $\mathcal{N}=1$, and the branes can fractionalize to $N$ fractional D4 branes (by separating them along the $R$ direction), each of which is described by pure $\mathcal{N}=1$ $SU(2)$ gauge theory at vanishing Chern-Simons level. On a different branch of moduli space, we can move the D4's away from the orbifold locus. If we keep them on top of the orientifold (at  $\text{Re}(\Phi_i)=0$), the gauge theory is symplectic, and has emergent 3d $\mathcal{N}=4$ at low energies. For a generic vev, away from the orbifold-orientifold singularity, the gauge theory is 3d $\mathcal{N}=8$ with gauge group $U(N/6)$, corresponding to the dimensional reduction of the $N/6$ mobile D4 branes on $T^2$. As described in Sections  \ref{sec:f32} and \ref{sec:f33}, this classical moduli space is lifted by quantum effects.}
\label{mspacef}
\end{center}
\end{figure}

Armed with these, we can study the classical moduli space of the probe D4 brane theory, which is depicted schematically on Figure \ref{mspacef}. Starting from the most symmetric point just described, we can access different branches of moduli space. Fractional branes living in the orientifold-orbifold locus correspond to giving a vev to the antisymmetric tensor scalar $\mathbf{R}^{(0)}$, which break the group generically to $Sp(1)^{N/2}$ (to see this, notice that $\mathbf{R}^{(0)}$ is an antisymmetric matrix, which can be always recast in a block-diagonal form where each block is proportional to the antisymmetric form of a single $Sp(1)$ factor; in the most general case, there are $N/2$ such blocks). All scalars are made massive by this vev and the low-energy gauge theory is 3d $\mathcal{N}=1$ pure gauge $SU(2)$ theory. On the other hand, a vev to bifundamental scalars breaks the group to the 3d $\mathcal{N}=4$ theory describing D4 branes sitting on the orientifold, or fully to the 3d $\mathcal{N}=8$ of branes at a generic point in moduli space.

Before finishing this section, there is just one more quantity that needs to be computed to determine the low-energy theory fully. On the 3d $\mathcal{N}=1$ branch of moduli space, the low-energy gauge theory also admits Chern-Simons term for each of the $Sp(N/6)$ factors (in the $\mathcal{N}=4$ and $\mathcal{N}=8$ branches, this is forbidden by supersymmetry). The Chern-Simons level for each of the $Sp(N/6)$ has to be identical, since they are related to each other by a discrete $\mathbb{Z}_3$ permutation  gauge symmetry. So there is only one Chern-Simons level $k$ that needs computing. The microscopic way to do this, starting from the 5d Yang-Mills theory of the D4 branes, is to evaluate the one-loop contribution to the Chern-Simons term coming from the towers of massive KK modes, after the orientifold and orbifold actions are taken into account.  Concretely, $k$ can be obtained from a regularized sum over KK modes \cite{Redlich:1983dv},
\begin{equation}k^{\text{1-loop}}=\frac12\lim_{s\rightarrow0}\sum_{a\in\text{KK tower}}\frac{\text{sign}(m_{a})}{\vert m_a\vert^s}. \label{calcu3}\end{equation}
This expression can be derived alternatively by using the adiabatic expression for the $\eta$ invariant that is the anomaly theory of the five-dimensional fermions, as explained in the Appendix of \cite{Dierigl:2022reg}. One arrives again at \eq{calcu3}; indeed, interpreting the masses of the KK tower as eigenvalues of a self-adjoint operator, the expression on the right hand side of \eq{calcu3} is precisely the $\eta$ invariant \cite{Witten:2015aba}.

In our case, however, we can argue that $k=0$ without explicit calculation, simply because there is a (Pin$^+$) parity symmetry in the brane system. The Chern-Simons term is odd under any parity symmetry \cite{Horava:1996ma,Witten:2016cio,Freed:2019sco}. The local D4-O6 intersection actually preserves one such symmetry, which was actually discussed in Section \ref{sec:parDGKT}. It corresponds, in the M-theory description, to reflecting the M-theory circle, as well as the 1,8, and 9 directions. As described in that Section, this parity symmetry is only broken by the first orbifold action in \eq{trolo}, so it survives as an exact parity symmetry of the local brane system. The KK towers in \eq{calcu3} appear in pairs with KK masses of opposite sign, and they cancel out as a result. With this, we learn that $k=0$, and we have a complete description of the low-energy effective field theory of the D4 branes.

\subsection{Spontaneous SUSY breaking and attractive force}

Equipped with the low-energy effective field theory of the probe D4 branes in all  the branches of moduli space, we can now study whether SUSY is broken spontaneously and a nontrivial vacuum energy is generated in the worldvolume theory. As we saw in Section \ref{sec:brbr}, the classical vacuum energy of the worldvolume quantum field theory corresponds to $T-q$ -- the linear combination of the brane tension $T$ and three-form charge $q$ that enters the BPS bound. A supersymmetric brane, sitting at the BPS bound, has $T=q$ -- the vacuum energy vanishes exactly and this corresponds to an exact moduli space. Spontaneous SUSY breaking where the vacuum energy density becomes $v\neq0$ corresponds to a correction away from the BPS bound,
\begin{equation} T-q=v.\end{equation}
Hence, if $v\neq0$ the brane will not be BPS. Since $v$ depends on the moduli (in fact, its dependence on the radial AdS coordinate $R$ is fixed by conformal/AdS symmetry, as reviewed in Section \ref{sec:brbr}), nonzero $v$ corresponds to a non-zero force felt by the brane, and a scalar potential in the $R$ direction.

In the rest of this Section, we will describe how a non-zero $v$ is generated for all the branches of moduli space described in the previous Subsection. This is compatible with the fact that in a 3d $\mathcal{N}=1$ QFT, the moduli space is unprotected and can receive quantum corrections, as mentioned in Section \ref{sec:brbr}. However, the corrections will be of different nature depending on the location of D4-branes. As we showed in Section \ref{sec:map}, the low energy worldvolume EFT is 3d $\mathcal{N}=1$ if the D4-branes sit at the orientifold-orbifold locus; while it has enhanced supersymmetry in the deep IR if the branes are away from the orbifold locus. This implies that only in the former case we can identify quantum effects within the EFT that break supersymmetry and generate a non-zero force. Otherwise, for the cases with enhanced $\mathcal N=4$ or $\mathcal N=8$ in the IR, we will have to look for perturbative or non-perturbative corrections to the superpotential of the EFT.
Perturbative corrections must come from the ambient fluxes and the non-trivial warped metric, which can generate worldvolume couplings beyond those discussed in Section \ref{sec:ang}. While quite generic, the problem is that this backreaction is difficult to compute without explicit expressions, which are only available in a few cases and then to low-orders only (see e.g. \cite{Marchesano:2020qvg} and Section \ref{sec:f33} for more details). Since we are interested mostly in the qualitative question of whether a nonzero $v$ is generated, and not so much its value, what we will do is to focus on non-perturbative contributions which, although subleading in most cases, can be computed reliably with only the zeroth-order Calabi-Yau metric and are enough to establish a departure of the BPS bound.

\subsubsection{D4 branes on top of the orbifold-orientifold singularity} \label{sec:f32}
 For branes on top of the orbifold-orientifold singularity, a generic point in the moduli space spanned by $R$ (where all branes are fully fractionalized) has a low-energy worldvolume field theory  which is 3d $\mathcal{N}=1$ pure $SU(2)$ gauge theory, at Chern-Simons level $k=0$. This theory was argued to spontaneously break supersymmetry in \cite{Witten:1999ds}. The gaugino field develops a condensate $\langle \bar{\lambda}\lambda\rangle$, which is a pseudoscalar. 
The low energy effective field theory can be obtained from truncating  \eqref{effaction} to the case at hand, i.e. a pure SU(2) gauge theory,
\begin{equation}S=   \int d^3\xi \left(-T_{\text{brane}}\,g_s^{-1}\text{Vol}(\omega_2)\frac{\ell_{\rm AdS}}{2R} \vert \nabla R\vert^2-\frac{1}{2g^2_{YM}}\vert F\vert^2\right)\ , \quad g^2_{YM}=\frac{R}{\ell_{\text AdS}}\frac{8\pi^2g_s\alpha'^{1/2}}{\text{Vol}(\omega_2)}\ ,\label{SU2}\end{equation}
with a dimensionful gauge coupling parametrized by the scalar $R$.
For convenience, we can re-write it in terms of the canonically normalized scalar field $\Phi$, which relates to $R$ by
\beq
\Phi\equiv\sqrt{\text{Vol}(\omega_2)T_{\text{brane}}\,g_s^{-1}4\ell_{\rm AdS}R}\ ,
\eeq
leading to
\begin{equation}S=   \int d^3\xi \left(-\frac12 \vert \nabla \Phi\vert^2-\frac{1}{2g^2_{YM}}\vert F\vert^2\right)\ , \quad g^2_{YM}=\frac{\Phi^2}{2\ell_{\rm AdS}^2}\frac{(2\pi)^6g_s^2 \alpha'^3}{\text{Vol}(\omega_2)^2}\ .\label{SU2-2}\end{equation}
where we have used \eqref{TQ}. Since the gauge coupling is the only scale in the theory, the scalar potential generated at the quantum level is given by
\beq
V\sim g_{YM}^6=\hat\lambda^{1/2}\left(\frac{\sqrt{2} R}{\pi \alpha'}\right)^3=\hat \lambda \,\Phi^6\label{vppo} \ .
\eeq
where $\hat\lambda^{1/3}\equiv 2^5\pi^6g_s^2 \alpha'^3(\text{Vol}(\omega_2)\ell_{\rm AdS})^{-2} $  is dimensionless. This matches with the expectations from \eqref{brascapo} and \eqref{awoo}. Notice in particular the cubic dependence on $R$, or equivalently the sixth power on $\Phi$, as foretold in \eqref{brascapo} in Section \ref{sec:brbr}. The contribution \eq{vppo} constitutes an additional contribution to the worldvolume vacuum energy of the brane; in other words, we have a positive contribution to $T-q$ since $\lambda> 0$. We conclude that these moduli space directions get lifted, and that this branch of membranes has $T>q$, and therefore they do not satisfy the Weak Gravity Conjecture. The branes feel an attractive force towards the Poincar\'e Horizon.

Nothing special happens when two fractional branes branes in this branch of moduli space coincide, but when a multiple of three are coincident, additional chiral multiplets become massless, allowing for recombination of the fractional branes so that they can move away from the singularity. At these points, the worldvolume theory is a more complicated 3d $\mathcal{N}=1$ theory with $SU(2)^3$ gauge group and bifundamental fields with quartic interactions, as discussed in the previous Section.  Computing the vacuum energy at this point is a complicated endeavor, but following the argument in Section 2.2 of \cite{Witten:1999ds} suggests that the Witten index also vanishes in this case, so supersymmetry is probably broken in this gauge theory. Although it would be desirable to have a more conclusive argument, this matches the picture of stringy instantons that we will discuss in Subection \ref{sec:f33}.

\subsubsection{D4 branes away from the orbifold-orientifold singularity} \label{sec:f33}

When the D4 branes are away from the orbifold, the deep IR worldvolume theory is 3d $\mathcal{N}=4$ (on top of the orientifold) or $\mathcal{N}=8$ (generically), so IR effects can no longer generate a superpotential. However, the D4 brane sits in a Calabi-Yau with fluxes, and there are additional effects to those captured by the low-energy effective field theory.  Here we will focus on non-perturbative effects coming from Euclidean D2 branes wrapping 3-cycles in the geometry. They will be sufficient to show that the moduli space gets lifted and the branes are no longer BPS, even if they do not provide necessarily the leading contribution to the potential.

We will start by briefly reviewing some generalities of instantons in supersymmetric theories. The interested reader can find more details in e.g. \cite{Blumenhagen:2009qh}. Then we will particularize it to the case of DGKT and explain why we generically expect to have instanton corrections that contribute to the worldvolume potential, although they are difficult to compute without having the full backreacted bulk solution. For the sake of concreteness, we will look for explicit solutions that exist even if only considering the local CY geometry (without taking into account the backreaction from the fluxes) and provide a concrete solution at the end of the section based on an Euclidean D2-brane intersecting the orbifold singularity.

\subsubsection*{Generalities from instantons}
In a general theory (SUSY or not), an instanton is a saddle point of some sort of Euclidean path integral that provides subleading contributions to observables. An instanton provides a contribution of the form
\begin{equation} P\, e^{-S},\end{equation}
to some observable of interest (say, the vacuum energy $\langle 1\rangle$), where $S$ is the euclidean action of the instanton configuration and $P$ is some prefactor. To evaluate $P$, one must perform a one-loop calculation around the instanton background, diagonalizing the operator of quadratic fluctuations of the effective field theory around the background. In this calculation, zero modes of the instanton background must be treated separately \cite{Coleman:1978ae}. In a Lorentz-invariant theory in $d$ dimensions, any instanton will have at least $d$ bosonic universal zero modes, parametrizing the spacetime location of the instanton. They can be understood as ``Goldstone'' modes from the spontaneously broken translational symmetry\footnote{Notice that these zero modes are not quantum field in any sense; there is no Goldstone theorem in 0 dimensions. This is merely a statement about the integral around the instanton background.}.

In theories with fermion fields, instantons may also develop one or more fermion zero modes $\lambda$. Since $\int d\lambda =0$ for a fermionic Grassman variable, and $P$ involves a path integral over $\lambda$, a fermion zero mode can cause instanton effects to vanish. Nevertheless, one can recover the instanton effects by changing the observable being studied; for instance, if instead of $\langle 1\rangle$ we compute $\langle \lambda \rangle$, then the integral $\int d\lambda\, \lambda =1$ and the instanton will contribute to this new observable.

In supersymmetric field theories, some instanton fermion zero modes are always guaranteed to exist. Since the only configuration invariant under all supercharges is the vacuum, an instanton background must spontaneously break at least some of the supercharges $Q_\alpha$ of the theory. These spontaneously broken supercharges become fermionic zero modes. A generic instanton will have as many fermionic zero modes as the total number of supercharges of the theory under consideration. In some cases (such as 4d $\mathcal{N}=1$, or 3d $\mathcal{N}=2$ theories), there may be BPS instantons, preserved by a subset of the supercharges. These preserved supercharges do not lead to fermion zero modes and, furthermore, bosonic and fermionic zero modes must fall in multiplets of the unbroken supersymmetry. No such relationship between fermionic and bosonic zero modes exists for non-BPS instantons.

In a supersymmetric field theory, we are often interested in instanton corrections to the Lagrangian. Due to supersymmetry, the Lagrangian can often be found via integrals on appropriate superspace. For instance, in 4d $\mathcal{N}=1$ theories, the Kahler potential is obtained from integration over four fermionic coordinates, while the superpotential involves only two. These superspace integrals can be saturated by instantonic fermion zero modes. As a result, an instanton contributing to the Kahler potential must have exactly four fermion zero modes, while one contributing to the superpotential must have two (and is therefore BPS). 
In the context of 3d $\mathcal{N}=1$ supersymmetric field theories of interest to us here, all of the action is obtained from a real superpotential $\mathcal{W}$, where instantons contribute if they have exactly two fermion zero modes (there are no BPS instantons in 3d $\mathcal{N}=1$).  Therefore, we will be looking for instanton effects on the D4 brane worldvolume theory with the minimal amount of two fermion zero modes required by supersymmetry. 

We also want to remark that fermion zero modes are extremely fragile, and that one should expect them to be lifted generically unless there is a good reason for their presence. For instance, a ``mass'' term $m \int \lambda_1 \lambda_2$ in the instanton action will pair up and lift fermion zero modes $\lambda_1$ and $\lambda_2$, since the path integral
\begin{equation} \int d\lambda_1\, d\lambda_2\, e^{-m \lambda_1 \lambda_2}= \int d\lambda_1\, d\lambda_2\, (1+ m\lambda_1\lambda_2)\end{equation}
automatically generates a term that saturates the integrals. But the same is true e.g. of cubic or quartic couplings $\int \lambda_1 \lambda_2 \lambda_3 \lambda_4$ between fermion zero modes, or any other higher-order coupling.  Fermionic quartic terms in D-brane actions are  related to higher-derivative terms, so to fully guarantee that fermion zero modes remain in an action one must take into account all these corrections. Conversely, it is natural to expect that, barring symmetry reasons, fermion zero modes will be lifted by one effect or another.  Therefore, even if we find that some instanton solution exhibits more than two fermion zero modes at low energies, they might be lifted at higher order and contribute to the potential anyway. It is noteworthy that the four-fermion and higher order couplings in D-brane actions, related to higher-derivative terms, are not known. Thus, more generally (and absent a parity symmetry), every 3d $\mathcal{N}=1$ system in string theory is in a dangerous position where uncontrolled effects can lift the scalar potential.

\subsubsection*{General expectations for DGKT}

Let us now consider the particular case of DGKT, and the effect of Euclidean D2 branes wrapping 3-cycles in the geometry. In the case where they wrap an orientifold-even cycle, they are projected in. One easy way to see this is to notice that, as discussed in Section \ref{sec:revw}, the RR three-form $C_3$ has zero modes on orientifold-even three cycles, and therefore there should be Euclidean D2-brane instantons that couple to the resulting four-dimensional axion. It is also possible for an Euclidean D2 to end on a D4 brane on a generic position; the Euclidean D2 is then wrapping a 3-chain ending on the D4 brane worldvolume, and not a  3-cycle. From the point of view of the 3d $U(1)$ worldvolume gauge theory of the D4-brane reduced on $T^2$, the Euclidean D2 endpoint is a monopole-instanton, a UV effect not captured by the low-energy effective field theory. The effect of these must be taken into account if we are to understand the low-energy dynamics of the system and, in particular, whether the branes generate a potential or not.

\begin{figure}[htb!]
\begin{center}
\includegraphics[scale = 0.75]{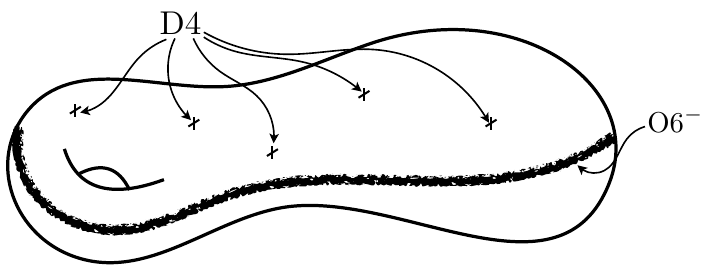}
\caption{Non-perturbative effects induced by D2 branes generate a repulsive force between the D4 branes on the configuration, similar to the effects in \cite{Khoze:2003yk}. The D4 branes spread out and end up in some energy-minimizing configuration on the Calabi-Yau (a generic Calabi Yau is depicted, rather than the orbifold one discussed in the text). Since the contributions to the scalar potential of the 3d $\mathcal{N}=1$ effective field theory are always positive, the non-perturbative potential is positive at the minimum, and the corresponding brane configuration sits above the BPS bound.}
\label{confinbr}
\end{center}
\end{figure}

In the following, we are going to consider Euclidean D2-branes stretching from the D4-brane to the orientifold. If these instantons have two  femionic zero modes (as we will argue below), they yield a contribution to the superpotential proportional to
\begin{equation} 
\label{inst-pot}\exp\left( - T_{\text{D2}} \text{Vol}_{\omega_2}\cdot d \right),\end{equation}
where $ \text{Vol}_{\omega_2}$ is the volume of the 2-cycle wrapped by the Euclidean D2 and D4 branes, and $d$ is the separation between the D4 and the O6 plane. This leads to a force dependent on $d$. Therefore, we  conclude that D4's in a generic position in the DGKT geometry feel a force coming from Euclidean D2 brane effects. This result is similar to the one of \cite{Brodie:1998bv,Aharony:1997ju,Aharony:1997bx}, where Euclidean D2 branes also generate a superpotential for D4 branes probing a 3d $\mathcal{N}=2$ geometry. Just like in that setup, the force is repulsive, since the instanton contributions to the superpotential are proportional to $e^{-d}$, where $d$ is the distance between branes, and the scalar potential is just the square of this. The D4 branes therefore repel each other (along the Calabi-Yau directions only), and their equilibrium configuration has all D4 branes dispersed throughout the Calabi-Yau (see Figure \ref{confinbr}). The exact equilibrium configuration depends on how many D4 branes we have, but in all cases the instantons will give a positive contribution to the energy, putting the system slightly above the BPS bound.  Therefore, the D4 branes at generic position do not obey the WGC for membranes, and combining with our analysis in Section \ref{sec:f32}, we find that actually, no D4 brane does.

The rest of this subsection is devoted to find an explicit configuration for such Euclidean D2 instanton, in order to check that it has the right number of zero modes to contribute to the superpotential as in \eqref{inst-pot}. Before we delve into the details, let us give an overall description of what we will find. Since DGKT is a 4d $\mathcal{N}=1$ solution,  the fully backreacted geometry only preserves a 3d $\mathcal{N}=1$ superalgebra, for which there are only two goldstini fermion zero modes for a generic instanton. In order to to find a concrete solution, we will first neglect backreaction effects and consider simply Euclidean D2-branes ending on D4-branes in a local Ricci-flat Calabi-Yau geometry (i.e., neglecting the effect of fluxes or warping). We will see that these instanton have too many fermion zero modes since they locally feel a 3d $\mathcal{N}=4$ geometry; we will find that they have four zero modes as corresponding to a 3d $\mathcal{N}=4$ BPS instanton. However, this analysis is merely an approximation. Since DGKT is a 4d $\mathcal{N}=1$ solution, the actual ten-dimensional metric is not Ricci flat. There are fluxes (Roman's mass, $H_3$ flux, and $F_4$ flux) everywhere throughout the Calabi-Yau, and they backreact on the geometry, as well as inducing warping. In particular, the backreacted metric is not Calabi-Yau; it is of more general $SU(3)\times SU(3)$ structure as explained in \cite{Marchesano:2020qvg}, where this backreaction is studied explicitly for a toroidal orbifold (different from the one we consider here). 
  Once corrections beyond the Calabi-Yau geometry are included and supersymmetry is broken to 4d $\mathcal{N}$=1,  we expect that the additional two fermion zero modes will be lifted by a combination of the fluxes and the geometry. \footnote{More concretely, fermion zero modes on the Euclidean D2 correspond to covariantly constant spinors in the D2 brane worldvolume with respect to the induced connection from the 10d ambient metric restricted to the brane worldvolume, and appropriately twisted by the fluxes (see e.g \cite{Simon:2011rw}).}
Hence, Euclidean D2 instantons will generically have just two fermion zero modes, and will contribute to the superpotential. This expectation is also supported by the \emph{fragility} of fermion zero modes (in the absence of a symmetry argument) discussed above. One does not even need a mass term to be generated in the worldvolume theory to lift them; but a four-fermion interaction (which leads to a classically gapless system in higher dimensions) can be enough to gap four fermion zero modes in one go \cite{Blumenhagen:2009qh,Dorey:2002ik}. 

Unfortunately, it is out of reach to study the fully backreacted geometry in DGKT to see explicitly how these additional zero modes get lifted. This is the same technical difficulty that makes impossible at the moment to fully uplift the DGKT vacuum to a 10d string theory solution. Nevertheless, it would be desirable to have,
 even as a proof of concept, an Euclidean D2 brane with just two fermion zero modes in the orbifold Calabi-Yau geometry, even before taking into account backreaction. Since away from the orbifold singularities the geometry is essentially toroidal, we need an Euclidean D2 brane intersecting an orbifold singularity to lift the fermion zero modes. At the end of this subsection, we will provide an example of such configuration.

\subsubsection*{On the quest of an instanton away from singularities}

\begin{figure}
\begin{center}
\includegraphics[scale = 0.65]{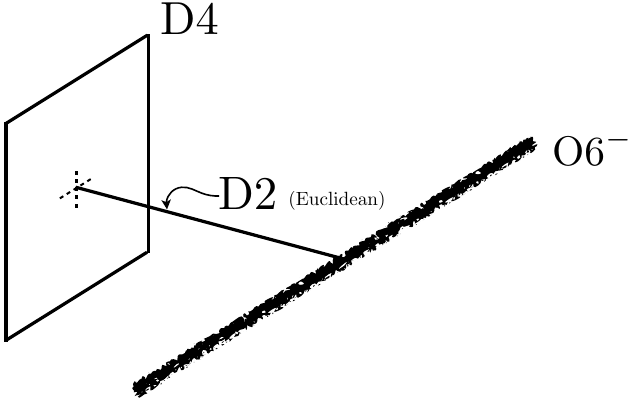}
\caption{A zoom-in on a D4 brane close to one of the components of the O6 plane in the DGKT geometry. There are Euclidean D2 brane instantons stretching from the D4 to the O6 that induce non-perturbative effects. In the leading approximation, where the local geometry is flat, the Euclidean D2 has four fermion zero modes, as described in the main text. Once the backreaction to the DGKT geometry is included, two fermion zero modes are lifted generically, and the Euclidean D2 induces a repulsive force between the O6 and D4, increasing the energy of the latter above the BPS bound. }
\label{zoomb}
\end{center}
\end{figure}

Consider first a D4 brane, possibly close to an orientifold (say, at a distance $d$), but far away from an orbifold singularity, as illustrated in Figure \ref{zoomb}. Consider then an Euclidean D2 brane wrapped as follows:

\begin{equation}
\begin{array}{c|cccccccccc}
\text{Brane}&T&X_1&X_2&R&X_4&X_5&X_6&X_7&X_8&X_9\\\hline
D4& \textrm{--}&\textrm{--}&\textrm{--}&\times&\textrm{--}&\textrm{--}&\times&\times&\times&\times\\
O6& \textrm{--}&\textrm{--}&\textrm{--}&\textrm{--}&\textrm{--}&\times&\textrm{--}&\times&\textrm{--}&\times\\
D2& \times&\times&\times&\times&\textrm{--}&\textrm{--}&\times&\textrm{--}&\times&\times\end{array}
\label{geom10}\end{equation}

The Euclidean D2 connects the D4 brane with its orientifold image. Although this Euclidean D2 brane is a legitimate, stringy euclidean instanton of finite action, it does not generate a scalar potential in the approximation just described. The reason is that, by zooming in close to the $O6$ and therefore neglecting effects from the orientifolds or the Calabi-Yau geometry, this system has 3d $\mathcal{N}=4$ supersymmetry; the Euclidean D2 brane under consideration is in fact a BPS 3d $\mathcal{N}=4$ instanton. To generate a scalar potential, one needs two fermion zero modes or less; $\mathcal{N}=4$ instantons typically have at least four (for BPS) or eight (for generic) fermion zero modes, so they will not contribute to the superpotential. We will nevertheless compute how many fermion zero modes are there for this instanton, since the calculation will be useful later.

 To do this, first consider an infinite D2, stretched as in \eq{geom10} and ignore the D4 brane. Since the D2 is a BPS brane, if it was in ten-dimensional flat space with no orientifolds, D4's, etc. the Euclidean D2 brane would have 16 fermion zero modes, corresponding to the sixteen spontaneously broken supersymmetries. The O6 projection on the D2 brane Chan-Paton factors is orthogonal, and therefore, for a single D2 brane, there is no action on Chan-Paton factors. The sole effect of the O6 plane is imposing a reflection on the 579 directions. This projects out half the fermions of the D2 brane, so eight fermion zero modes are preserved\footnote{The easiest way to see this is to T-dualize along the 589 directions, to a type I picture where the D2 becomes a D1 brane. The orientifold projection on the D1 projects haf the fermion zero modes, since the type IIB $\Omega$ action acts on the supercharges nontrivially \cite{Tachikawa:2018njr}. One may also S-dualize to heterotic, which famously has 8 fermionic fields (superpartners to the right-moving spacetime bosonic coordinates).}. This is what one expects for a BPS instanton in the 3d $\mathcal{N}=8$ background of the orientifold, and indeed, the D2 arranged in \eq{geom10} is mutually BPS with the O6 as they have a number of relative Neumann-Dirichlet directions which is a multiple of four \cite{polchinski1998string}.

\begin{figure}[htb!]
\begin{center}
\includegraphics[scale = 0.75]{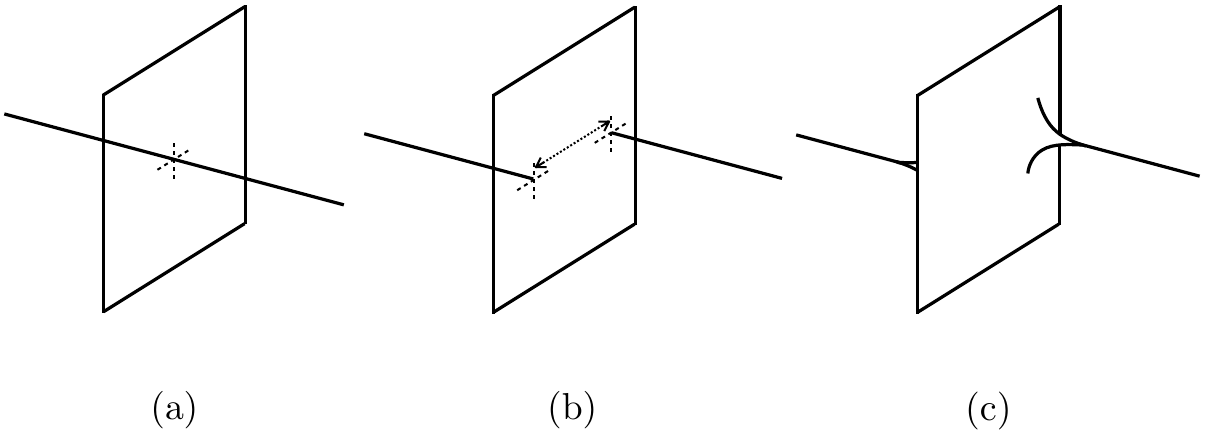}
\caption{(a) An infinite D2 intersecting a D4 brane transversally. There are bifundamental fileds at the intersection. (b) Giving a vev to the bifundamental scalars described in panel (a) corresponds to splitting the D2 into two semi-infinite D2's, each of which ends in the D4. There is a D4 worldvolume monopole at the endpoint. (c) The backreacted picture of the D2 endpoint is that of a ``bion'', where the D2 worldvolume pulls on the D4 and they recombine. This has the effect of lifting several fermion zero modes as described in the main text. }
\label{bion}
\end{center}
\end{figure}

The above story is for an infinite Euclidean D2, which has infinite action and is therefore not interesting to us. We are interested in an Euclidean D2 which ends on the D4 brane worldvolume. To understand this effect, consider now an Euclidean D2 interseting the D4, far away from the O6 for the time being\footnote{We are indebted to A. Uranga for explaining this approach to us.}. There are, as usual, bifundamental fields charged under the D2 and D4 gauge groups on the intersection. Giving a vev to the bifundamental scalar corresponds to ``breaking up'' the D2 brane into two pieces that end on the D4 \cite{Diaconescu:1996rk,Brodie:1998bv,Johnson:2000ch,Khoze:2003yk,Moore:2014gua}, see Figure \ref{bion}. The effect of this vev is that Euclidean D2 deformations transverse to the D4 become massive (it is not possible to move the D2 ``away'' from the D4, since it is ending on it). There is also a coupling between the bifundamental scalar, the bifundamental fermion at the intersection, and bulk D2 fermions, so giving a vev to the scalar also switches on Neumann boundary conditions for some of the bulk D2 fermions, which therefore lose their fermion zero modes. Therefore, a finite Euclidean D2 brane has less fermion zero modes than its infinite counterpart. The physical interpretation of this is that the D2 ``pulls'' on the D4,  recombining with it and forming a ``bion'' configuration   (see rightmost panel of Figure \ref{bion}). To find out how many fermion zero modes are lifted, we can T-dualize the D2 ending on D4 to a D1 ending on D3, a configuration which famously preserves eight supercharges and has eight fermion zero modes \cite{Khoze:2003yk} due to the bion effect. The lifting of exactly four fermion zero modes can also be understood from the fact that the D1-D3 intersection preserves eight supercharges, so zero modes (bosonic and fermionic) must fall in representations of the unbroken supersymmetry. Since the bion effect lifted all bosonic normal directions to the D1 and D3, the corresponding number of fermion zero modes must be lifted, too\footnote{Note that upon T-dualizing back to D2-D4, some of the lifted normal directions in the D1-D3 correspond to lifted Wilson lines in the D2 worldvolume. The interpretation is that the boundary conditions on the D2 gauge field imposed by the D4 are such that these fermion zero modes are absent.}. Since breaking up the D1 or D2 into pieces preserves supersymmetry, we should think of a D1 that does not end on D3 or a D2 that does not end on D4 as effectively a ``two-instanton'' configuration, with the ending D1 or D3 being the ``single instanton'' configuration.  Hence, the number of fermion zero modes on an ending D1 or D2 is half of those on an infinite one.  Furthermore, the D4 breaks the supersymmetries to 3d $\mathcal{N}=8$, and the D2 is also BPS; the generic number of fermion zero modes for a BPS instanton is in this case eight.

We can now re-include the effect of the orientifold projection, which, as before, halves the number of fermion zero modes, down to four. We can again cross-check this by T-dualizing in the 589 directions, to type I, where the D4 becomes a D5 and the D2 becomes a D1. D1 has eight fermion zero modes; since the D1 can end on the D5, four of these are lifted by bion effects, leading again to the four fermion zero modes of a 3d $\mathcal{N}=4$ BPS instanton -- indeed, the D2 is mutually BPS to the O6-D4 system which preserves 3d $\mathcal{N}=4$.

One may wonder if this conclusion may be altered by the presence of Roman's mass. Unlike other fluxes, or the overall CY geometry, the Roman's mass background behaves as a ten-dimensional cosmological constant, and it does not dilute as the Calabi-Yau space is taken to be arbitrarily large. If we did not have the orientifold, the system would be just a stack of D4 branes in massive IIA string theory, which preserves eight supercharges; as a result no scalar potential can be generated, and a generic instanton  (such as the Euclidean D2 brane) has eight fermionic zero modes. Including the O6 projects the system down to 4d $\mathcal{N}=1$ (four supercharges)\footnote{This is perhaps clearer if one imagines adding a $D8$ brane transverse to the 6 direction, which sources the Roman's mass. The transverse directions between D4 and D8 are 3789, while those between O6 and O8 are 5679. The fermion lift of parity transformations along these directions commute with each other, and they have a single common eigenvector, leading to 4d $\mathcal{N}=1$ \cite{polchinski1998string}.}.  One can check that the Euclidean D2 brane is then non-BPS due to the additional breaking of supercharges caused by Roman's mass, and will have four fermion zero modes -- again, too much to contribute to the scalar potential.  

To sum up the above analysis, where we have focused in studying the local geometry of the system, ignoring the global Calabi-Yau geometry and the subleading backreaction caused by the fluxes and warping, we have not found an instanton that contributes to the D4 tension. As anticipated, the reason is that we are working on an orbifold Calabi-Yau and therefore, ignoring backreaction and away from the fixed loci, the Calabi-Yau metric is actually Riemann-flat. The local O6-D4 geometry preserves 3d $\mathcal{N}=4$, and no corrections to the superpotential can develop. 
In order to lift the additional zero modes, we will need to consider an euclidean D2-brane intersecting an orbifold singularity, as we study next.

\subsubsection*{Explicit configuration at the orbifold singularity}

\begin{figure}[htb!]
\begin{center}
\includegraphics[scale = 0.7]{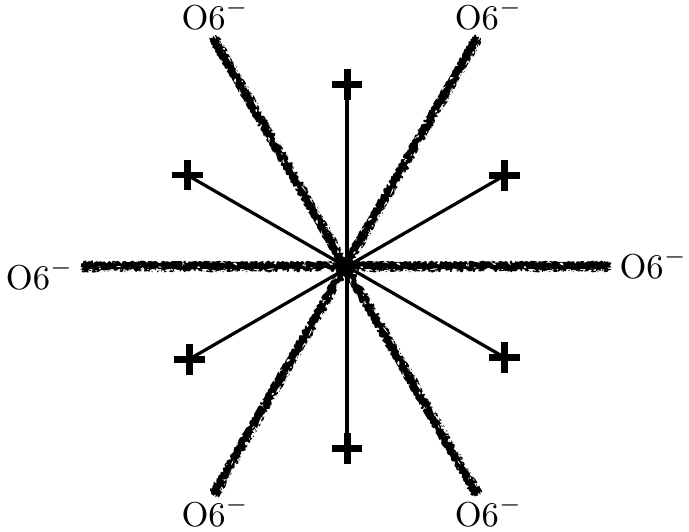}
\caption{An Euclidean D2 brane stretching between a D4 near an orbifold-orientifold singularity and its images. Here, just the $x_6=\text{Re}(z_2)$ and  $x_7=\text{Im}(z_2)$ coordinates are depicted; the D4 brane location and its orbifold images are marked by crosses, while the O6 and its orbifold images correspond to thick lines. The thin lines stretching from the D4 to the O6 singularity correspond to the Euclidean D2 brane and its orbifold images. This configuration locally sees the 3d $\mathcal{N}=1$ geometry, and as discussed in the main text, it leads to the kind of instanton with two fermion zero modes that we seek.}
\label{hex-1}
\end{center}
\end{figure}

Consider now a D4 brane in the vicinity of an orientifold-orbifold singularity, as illustrated in Figure \ref{hex-1}. This actually describes a D4 in general position in the Calabi-Yau, since the global geometry is just a flat quotient away from these singularities. The orbifold-orientifold group at the singularity is $\mathbb{Z}_6$, so there are pure orbifold images of the D4 (corresponding to elements of even order) and three orbifold-orientifold singularities (corresponding to orbifold elements of odd order). We will consider an Euclidean D2 brane stretching between the D4 and its orbifold-orientifold image of order three, as illustrated in Figure \ref{hex-1}. In general, we would need to consider such D2 branes connecting the D4 with its images for all orbifold-orientifold singularities in the global geometry; but studying just one will suffice for our purposes. 

Let us consider first the effect of the  O6$^-$ on the D2 fermions. In the configuration of Figure \ref{hex-1}, the action of the orientifolds maps the D2 brane images to each other, in three pairs of two. Each pair can be analyzed separately and each is identical to one copy of the D2-O6$^-$ system that we just discussed. Therefore, each pair leads to a locally 3d $\mathcal{N}=4$ system, for which the Euclidean D2 constitutes a BPS instanton (including Romans' mass does not change this counting, as explained above) with four worldvolume fermion zero modes once the bion effect is taken into account. We can also see this directly, without appealing to supersymmetry: The Euclidean D2 brane is wrapped along the 457 directions, as depicted in \eq{geom10}.  Since the worldvolume theory of any D-brane is the dimensional reduction of 10d $\mathcal{N}=1$ SYM, there is an action of the 9d Clifford algebra (the largest Clifford algebra that acts on 10d chiral fermions) on the fermions. The matrices $\Gamma_{4,5,7}$ represent the Clifford algebra associated to the worldvolume directions of the brane, while the rest represent the action on the normal bundle. There is no additional action on Chan-Paton factors of each pair since the fermions are neutral under the D2 brane gauge group, which is $U(1)$ (since the projection is orthogonal, we do not need to take a pair of branes). The fermion zero modes are those spinors that satisfy 
\begin{equation} \Gamma_5 \Gamma_7 \Gamma_9 \lambda=\lambda,\label{ori55}\end{equation}
an equation that has eight fermion zero modes as solution. These correspond precisely to the eight fermion zero modes on an infinite D2; if the D2 ends on the D4, four of these are lifted by the bion effect. Finally, since there are three pairs of orientifold-identified D2-brane stacks, we would get a total of 12 fermion zero modes. However, imposing the orbifold projection permutes each D2 brane pair into the next, and acts as a rotation by $120^\circ$ on the $T^2$ parametrized by $z_1$. To find a fermion zero mode of the orbifold action, we need to combine the eigenfuctions of the geometric part of the orbifold with those of the action on Chan-Paton factors, which is a three-dimensional permutation matrix acting on the three branes. Combining with the geometric part, we find a total of four bulk fermionic zero modes living on the worldvolume of the D2's; the only effect of the orbifold is to identify the fermion zero modes of each orientifold-identified D-brane pair separately with each other.

In addition to these fermionic worldvolume modes,  
there will also be bifundamental, massless charged fields located at the brane intersection for any pair of intersecting D2-branes. The D2 branes intersect on a single, real plane, so we can obtain the spectrum from e.g. the dimensional reduction on $T^2$ of the intersecting D4-brane models of \cite{Aldazabal:2000dg}. The result is that there are two complex chiral bifundamental fermion fields of each chirality living in the two-dimensional intersection of the D2 branes; these yield a total of eight fermion zero modes living at the intersection per each pair of D2-brane images identified by the orientifold. The orientifold action acts as charge conjugation on these fields, and projects down the number of fermion zero modes on the intersection down to four. This yields a total of 12 fermion zero modes when taking into account the three intersecting copies of the D2-O6$^-$system. However, the orbifold projection again reduces this to only four fermionic zero modes,  due to a similar effect to the one discussed above. 
There are also bifundamental charged scalars that have a similar fate. 

\begin{figure}[htb!]
\begin{center}
\includegraphics[scale = 0.6]{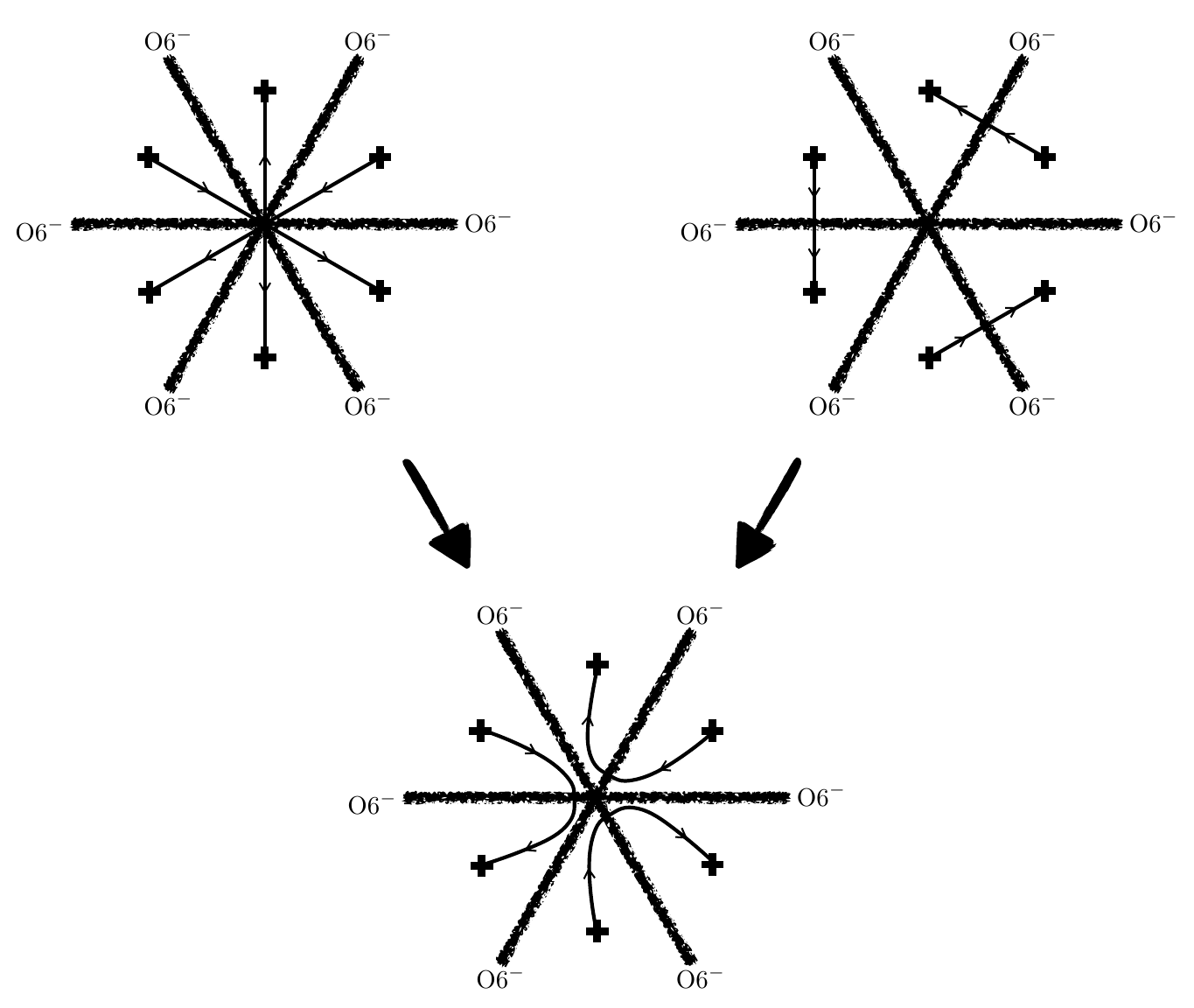}
\caption{The Euclidean D2 brane configuration where all the brains intersect at the singularity is a saddle point, but not a minimum. It can lower its energy via brane recombination as depicted. Correspondingly, there is a tachyonic field at the intersection, coming from perturbative strings quantized at an angle on a plane, whose vev is the deformation parameter. The endpoint of the configuration, depicted on the bottom panel, has no fermionic zero modes other than the minimal two required by supersymmetry. In this configuration, the D2 branes (whose orientation in the covering space is depicted by arrows) feel an attractive force towards their orbifold images, resulting in bending of their worldvolumes. Due to this attractive force, the configuration in the bottom panel also has less action than that on the top-right panel, which depicts three copies of the configuration in Figure \ref{zoomb}; this would be the minimum action configuration if only the tension term for the action of each Euclidean D2 brane is considered.}
\label{hex-2}
\end{center}
\end{figure}

To summarize, we have four fermion zero modes living at the intersection, as well as another four living on the bulk of the D2-brane worldvolume. To obtain the promised instanton that contributes to the superpotential, we must perform one last step.  The configuration of Euclidean D2 branes depicted in Figure \ref{hex-1} is not a minimum of the action -- only a saddle point--. The D2's intersect their orbifold images at an angle on a plane, and as is well-known, in these configurations the bifundamental scalar discussed above becomes a tachyon \cite{Epple:2003xt}. Giving it a vev corresponds to a brane recombination effect, as shown in Figure \ref{hex-2}. When the tachyon obtains a vev, all fermion zero modes living at the intersection become massive due to tachyon couplings.\footnote{We thank A. Uranga for discussions on this point.} Moreover, the existence of trilinear couplings involving each of the localized fermions, the tachyon and a bulk fermion living on the D2 worldvolume, generates a mass for two of the bulk D2 fermion zero modes\footnote{This counting is easier to understand if taking the bion effect as the final step. Notice that before taking into account the bion effect, there would be eight bulk femion zero modes and four localized fermionic zero modes in the intersections. The trilinear couplings would generate a mass for four of these eight bulk fermions. The bion effect then reduces the number of zero modes to two.}. The resulting configuration -- the endpoint of the tachyon condensation process --, seen at the bottom panel of Figure \ref{hex-2}, has exactly only the two fermion zero modes required by supersymmetry, and provides the superpotential contribution we were looking for. 

The reason why the minimum action configuration is that of the bottom panel of Figure \ref{hex-2} is that each of the Euclidean D2 branes feels the presence of its orbifold images; since the orientations of the Euclidean D2 branes are as depicted, each brane feels an attractive force towards its images, as would a brane-antibrane pair in empty flat space; the point is that, on top of the dilaton-gravitational force between them (which is always attractive), each line element $d\vec{l}$ of D2 brane feels a force from any other element $d\vec{l}'$ which is inversely proportional to a power of their distance and whose sign is controlled by the dot product $d\vec{l}\cdot d\vec{l}'$; this quantity is negative for the orientations of the D2 branes depicted in Figure \ref{hex-2}. If there were no forces, gravitational or gauge, between the branes, each of them would minimize their action by relaxing to a configuration where the D2 branes are straight, as in the top right panel of Figure \ref{hex-2}; such configuration is locally identical to that discussed earlier and would have additional fermion zero modes instead. However, due to the attractive forces just discussed, the branes will bend towards each other, as in the bottom panel of Figure \ref{hex-2}. We will not attempt to find the minimum configuration, which would be quite convoluted due to the geometry, the fact that the long-range field of each D2 brane is somewhat screened by the magnetic monopoles at their endpoints, and the non-trivial dilaton and gravity profiles; but the above arguments strongly suggest that one exists. On such a generic configuration, where the brane is wrapping a generic curve hanging from the two D4's at its endpoints, there are no additional fermion zero modes, other than the minimum two required by supersymmetry. 

The result makes intuitive sense: although the orbifold Calabi-Yau seems almost Riemann-flat, and hence locally higher-supersymmetric, when we put together the effects of both orbifold and orientifold, which are enough to break supersymmetry to 3d $\mathcal{N}=1$, we find effects that generate a superpotential. This implies that the D4 brane is no longer BPS and it will not satisfy the Weak Gravity Conjecture; it will feel an attractive force as discussed in Section \ref{sec:brbr}.

\begin{figure}[htb!]
\begin{center}
\includegraphics[scale = 0.4]{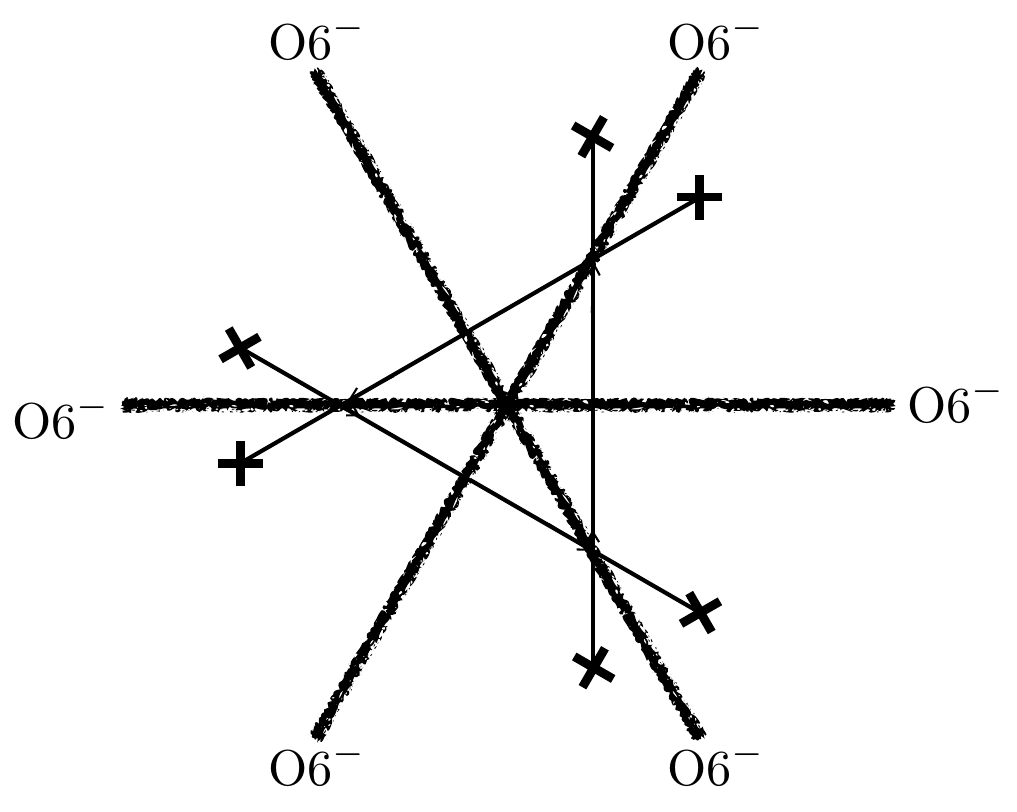}
\caption{More generic configuration where the D4 branes are not exactly halfway between the O6 planes. In this case, the Euclidean D2 branes do not intersect exactly at the orbifold-orientifold locus, but rather on top of the orientifold at an angle. The effect of the orbifold projection is however similar, and the recombination to three disconnected saddles proceeds in the same way.}
\label{hex-3}
\end{center}
\end{figure}

The analysis so far was carried out for D4 branes sitting at the same distance of the O6$^-$ planes on either side. More generally, the D4 brane will be closer to one O6$^-$ than the other; the resulting configuration for the Euclidean D2 branes is depicted in Figure \ref{hex-3}. In spite of the different picture, the analysis is exactly the same; the orientifolds still project the branes in pairs of two, and the orbifold again merely permutes one Euclidean D2 brane pair to another. Similarly, the Euclidean D2 branes recombine to a configuration like that on the bottom panel of Figure \ref{hex-2}. The continuity of non-perturbative effects under deformations of the underlying configuration is reminiscent of \cite{Garcia-Etxebarria:2007fvo,Garcia-Etxebarria:2008mpu}, which discussed a similar phenomenon in the  4d $\mathcal{N}=1$ context.

Additionally, there could be other Euclidean brane instantons that contribute to the D4 tension. For instance, there could be worldsheet instantons stretching between the orientifold and the D4 brane. We have not explored these effects, and mention them just so that it is clear that, as one would naturally expect in a 4d $\mathcal{N}=1$ setting, many quantities of interest receive corrections due to the small amount of supersymmetry.
 
Finally, we should entertain the possibility that the various non-perturbative (as well as potential perturbative) corrections to the superpotential entertained in this Section may all have different signs (which depend on uncalculated 1-loop instanton prefactors), so that they might cancel out and generate a supersymmetric contribution in a particular locus of the brane moduli space. Although we cannot exclude it, we view this possibility as unlikely: The (real) F-terms are sections of the tangent bundle of the moduli space, and a generic section of a bundle with vanishing first Stiefel-Whitney class does not have any zeroes. 

\subsubsection{Other brane configurations}\label{sec:obc}

Given the fact that the above wrapped D4-branes do not satisfy the Weak Gravity Conjecture, one may wonder if there could be other brane configurations that do. This seems very unlikely, on account of the generic character of corrections to the superpotential in 3d $\mathcal{N}=1$ discussed in the previous Section, but let us nevertheless discuss the possibilities. 

To satisfy the WGC, it is enough to find (for each charge direction) at least one membrane satisfying the bound on the charge-to-tension ratio. We have already discarded brane configurations with just D4 branes. The next option is to look for other brane configurations that have $C_4$ charge, like e.g. D6-branes with worldvolume flux. However, these branes are generically non-BPS already at the classical level. The supersymmetric conditions for wrapping D-branes were  studied in \cite{Aharony:2007du}, concluding that, in a generic situation, the D4-branes were the only supersymmetric branes at the classical level. The D6-branes could also become BPS in principle if adding a worldvolume flux; however, the supersymmetry condition relates the induced metric on the wrapping torus and the worldvolume flux in a way which is not typically satisfied for integer values of the flux.  This means that, already at the classical level, they will have a tension bigger than the charge (because of the BPS bound), and they will be self-attractive. Considering quantum corrections will only reinforce this, as their contribution to the superpotential is positive. Furthermore, any corrections will be subleading to the classical effects for weak enough dilaton, which can always be achieved by increasing the 4-form flux $N$ in the DKGT solution. 

Recently, a more in-depth analysis of the possible BPS branes at the classical level that can exist in DGKT-like vacua was carried out in \cite{Marchesano:2022rpr}. There, new classically-BPS branes associated to D8 branes, with or without worldvolume fluxes, were described. 

A classically BPS D8 brane will change Romans' mass and may also change the $F_4$ flux, if it carries worldvolume fluxes corresponding to D4 charge. In either case, the D8 must be attached to D6 branes on one side, to compensate for the change in the O6 tadpole resulting from the change in Romans' mass. As DGKT vacua with D6 branes are more energetic than those without, these BPS branes only exist in a DGKT vacuum that originally contains D6-branes. This is why we have not considered them in more detail, since they cannot provide a general candidate to satisfy the WGC for DGKT vacua without D6 branes, which are the focus of our paper. Another possibility in  \cite{Marchesano:2022rpr} consists of anti-D8-branes with wolrdvolume fluxes, which could be BPS even if the vacuum has no D6-branes. However, they suffer from the same problem discussed above for D6-branes; the supersymmetric conditions can only be solved for very specific values of the fluxes, and therefore these branes are not BPS in a generic DGKT vacuum. 

Taking stock of the above, the D4's seem to be the only BPS branes at the classical level for any choice of the bulk fluxes. Even in the particular instances in which the other BPS branes exist,  
we expect them to receive corrections leading to $T>q$ just like the D4 branes, and from similar effects (Euclidean D2 branes starting on the orientifold can easily end on the D8 or D6 branes, just like they do on D4's). In any case, none of these branes could provide an object satisfying the WGC along the particular direction of the charge lattice associated purely to the $F_4$ flux, since they have additional charges along other directions.

Finally, although our analysis focused on the original DGKT orbifold Calabi-Yau, the arguments are very general, and we expect them to hold for different Calabi-Yau's. Actually, for a general Calabi-Yau, the Euclidean D2 branes of Section \ref{sec:f33} will locally see 4d $\mathcal{N}=1$ geometry, and we expect that the lifting of fermion zero modes is apparent even before taking into account backreaction. From this perspective, the orbifold Calabi-Yau geometries seem harder to analyze than the general Calabi-Yau.

\subsection{Interpretation of results and tension with the WGC\label{sec:interpretationWGC}}

We got ourselves an interesting puzzle. On one hand, we have seen that, assuming that the supergravity DGKT solution is indeed a good low energy description of a 4d $N=1$ AdS quantum gravity vacuum, we have obtained that the wrapping D4-branes that mediate flux jumps feel an attractive force towards the origin of the Poincare horizon. In global coordinates, it means that non-perturbative bubbles that could discharge the flux actually shrink after nucleation, so that they do not describe a dynamical decay. This is consistent with the fact that the vacuum is supposed to be supersymmetric and, therefore, stable; providing a nice consistency check of the supergravity solution. However, it does not give us information about the existence or absence of a string theory 10d uplift of the solution, since we are assuming from the start that the 4d supergravity action used in the DGKT proposal is a good approximation of the low energy description.

On the other hand, we would have expected by the Weak Gravity Conjecture that these branes should feel a repulsive (rather than attractive) force, yielding an instability of the vacuum. As mentioned in Section \ref{sec:brbr}, given a quantum gravity theory with a  massless U(1) gauge field, the Weak Gravity Conjecture requires the existence of at least one electrically charged state with a charge to mass ratio in Planck units bigger or equal than the charge to mass ratio of an extremal black hole in that theory. This is one of the best understood and more solid Swampland criteria; the evidence ranks from plethora of top-down string theory data (see e.g. \cite{Heidenreich:2016aqi,Lee:2018urn,Gendler:2020dfp,Klaewer:2020lfg,Cota:2022yjw,Cota:2022maf,FierroCota:2023bsp,Gendler:2022ztv}) to bottom-up arguments based on black holes \cite{ArkaniHamed:2006dz,Hamada:2021yxy} and causality of scattering amplitudes \cite{Hamada:2018dde,Cheung:2018cwt,Bellazzini:2019xts,Arkani-Hamed:2021ajd,Henriksson:2022oeu}, as well as derivations in string perturbative theory \cite{Heidenreich:2016aqi,Heidenreich:2024dmr} and holography \cite{Montero:2016tif,Montero:2018fns} (see \cite{Harlow:2022ich,Rudelius:2024mhq} for reviews). If applied to the top-form gauge fields dual to the internal fluxes that support the AdS solution, it seems to imply the existence of a codimension-1 object (i.e. a membrane in 4d) which is magnetically charged under the internal flux and whose tension is smaller or equal than its charge in Planck units, such that it is self-repulsive (i.e. the gravitational force acts weaker or equal than the gauge force). This implies that there should be a brane mediating flux jumps that either feels no force, or this force is repulsive, yielding an instability of the vacuum. This is the argument that was used in \cite{Ooguri:2016pdq} to propose that any non-supersymmetric AdS vacuum supported by fluxes should be at best metastable. We have seen that, in the absence of parity symmetries, this argument can be extended to 4d N=1 AdS vacua, since the superpotential of the corresponding brane is not protected and, thus, one cannot satisfy the WGC by saturating the equality; rather, there should be at least one brane feeling a repulsive force, which makes the vacuum unstable and, therefore, inconsistent as a SUSY solution.

In conclusion, we have explicitly shown that the wrapped D4-branes charged under the $F_4$ flux in the DGKT solution feel an attractive force and, therefore, violate the Weak Gravity Conjecture applied to membranes. Hence, there are only two possible resolutions. Either the DGKT vacuum is not a consistent solution of quantum gravity, or the Weak Gravity Conjecture applied to membranes must be revisited since this would present a counterexample. Further research is needed to determine which resolution is correct; it is not a priori clear what the right answer is since the Weak Gravity Conjecture for membranes in 4d is much less understood than for particles. If we could show that DGKT presents a violation of the WGC for particles, then the story would be much more clear, signaling bad news for DGKT. However, the situation is different for codimension-1 objects, since the backreaction is so strong that breaks the asymptotic structure of the vacuum and does not allow for the existence of low codimension black branes.  Hence, the huge amount of evidence for the WGC that we referred above, does not necessarily apply for low codimension states. For instance, since we do not have black brane solutions, they cannot be used to fix the exact order one factors in the WGC inequality. Instead, they are fixed by requiring a balance of forces, which is a different interpretation of the WGC often known as the Repulsive Force Condition (RFC)  \cite{Palti:2017elp,Heidenreich:2019zkl}. Both interpretations coincide in the absence of scalar forces, but can differ otherwise. We want to note, though, that the differences between the WGC and the RFC typically become negligible at weak coupling, which is precisely the case of interest in DGKT (at small $g_s$, the D4-brane becomes weakly coupled). Moreover, the tension and charge of a membrane in 4d are RG-flow dependent quantities, and their value depends on the distance from the object at which they are evaluated (see \cite{Lanza:2020qmt} for a discussion of the WGC for low codimension objects taking into account the backreaction of scalar fields driving the RG flow). In any case, it is expected that some version of the WGC should survive for low codimension objects (since they are connected by dimensional reduction to higher codimension ones), but we cannot discard the possibility that subtleties might arise that could be crucial here. Additional subtleties could also arise from the fact that we are applying the WGC in AdS space rather than in flat space; see \cite{Nakayama:2015hga,Montero:2016tif,Montero:2018fns,Aharony:2021mpc,Palti:2022unw,Sharon:2023drx,Andriolo:2022hax,Orlando:2023ljh} for proposals about how to define the WGC for particles in AdS space from the perspective of the CFT dual.

Moreover, with the techniques above we are only able to study stacks of a number $k$ of D4 branes in the DGKT geometry. Since we work with probe branes in a weakly curved background, we need to take $k\ll N$. One could then entertain the possibility that there is a qualitative change in dynamics for a stack of $k\sim N$ D4 branes, and that such a stack (including appropriate backreaction in the geometry) turns out to be supersymmetric. In this case, we would say that the WGC is preserved, although only with a sublattice of order $\sim N$. It would be interesting to explore this possibility further, by  e.g. investigating bubbles of nothing dressed by $F_4$ flux in the DGKT scenario, but we do not think it can happen. Suppose that one such solution with D4-brane charge of order $N$ existed; then we could stack together $m$ copies of this supersymmetric domain wall, to produce a new copy of the DGKT model with total $F_4$ flux of order $m N$. For $m\gg N$, the original domain wall is a small perturbation on a weakly curved geometry and should fall within the purview of our analysis, concluding that the domain wall is not BPS.

One could also wonder if there can be other branes satisfying the WGC in the DGKT solution beyond the ones we studied in this paper. We discussed in Section \ref{sec:obc} all possible brane configurations (to the best of our knowledge) that could play this role. Unless we missed something, there is no candidate that could satisfy the WGC.

Therefore, to the best of our knowledge, this is the first in the literature  that a supersymmetric AdS solution violating the WGC for membranes is found\footnote{There is another potential candidate for a 4d $\mathcal{N}=1$ vacuum \cite{Gaiotto:2009yz} that might experience a similar fate than DGKT, as we elaborate in the conclusions.}. There is a somewhat similar situation in the holographic dual to the D1-D5 SCFT \cite{Seiberg:1999xz}, which is dual to IIB on $AdS_3\times S^3\times T^4$\footnote{We thank Juan Maldacena for bringing up this example.}. This holographic background is threaded by $N$ units of $F_3$ flux on the $S^3$ and $M$ units of $F_7$ flux on the $S^3\times T^4$. Relatedly, these fluxes can be discharged by D1 and $D5$ branes. This is a 3d $\mathcal{N}=4$ setup and, depending on the precise values of the moduli, a single D1 or D5 are not in general BPS, even at the classical level, due to the classical backreaction of the moduli profiles sourced by the brane.  However, in this setup, a bound state of $p$ D1 and $q$ D5 branes with $(p,q)$ proportional to $(M,N)$ is always BPS, and hence provides an exactly BPS membrane, so that the WGC is still satisfied, with an index of order $N$, and the arguments of \cite{Ooguri:2016pdq} still hold. By contrast, in DGKT we find that branes in all directions of the charge lattice are lifted, including those completely aligned with the background fluxes.  

Therefore, our result for DGKT is interesting by itself regardless of the WGC. If the DGKT solution exists, this could be an important hint towards constructing the CFT dual. The existence of only self-attractive non-BPS branes is a feature which is not present so far  in any of the known holographic (SUSY) examples, for better or worse. In an upcoming work \cite{fien}, we will also describe the singularity in which one should place the stack of branes whose near horizon geometry should (a priori) reproduce the DGKT AdS solution. 

We hope that a better understanding of these attractive branes can help to determine the fate of DGKT as a quantum gravity solution. Either it indeed implies that DGKT is UV inconsistent (so that it is in the swampland), or it forces a modification of the WGC for fluxes, which would have tremendous implications regarding the stability of  AdS quantum gravity vacua with zero or four supercharges.

\section{General implications for 4d \texorpdfstring{$N=1$}{N=1} vacua and conclusions}\label{sec:s5}

In this paper we have pointed out and explored the tension between the membrane WGC and the DGKT scenario. To do this, we have taken the first steps in investigating the fate of the moduli space of the putative DGKT field theory dual at quantum level, showing via explicit computation that it gets lifted by quantum corrections and instanton effects. This is consistent with the fact that the DGKT vacuum does not preserve any parity symmetry that could protect the 3d $\mathcal{N}=1$ moduli space, and implies that the membranes charged under the bulk $F_4$-flux are self-attractive, violating this way the expectations from the WGC. If a membrane version of the WGC can be shown to hold, then (modulo the subtleties discussed in Section \ref{sec:interpretationWGC}), the DGKT scenario must be pathological in some way. 

Could the membrane WGC simply be false?  The low-codimension versions of the WGC have always been a bit apart from their higher-codimensional counterparts. This is because there are generically no black brane solutions with a horizon in low codimension, so there is no direct connection between WGC and black brane decay\footnote{In these cases, the WGC is defined as a repulsive force condition, since there is no extremality bound to compare with.}. Therefore, there could be subtleties regarding the exact WGC inequality that one must satisfy. On top of this, the proper definition of the WGC in AdS space may be subtler than in flat space. All these caveats were discussed in Section \ref{sec:interpretationWGC}. 

On the other hand, if the membrane WGC is false, there is no reason why we should not have seen this failure in more supersymmetric examples or at the classical level, where instead it seems to have been verified in many examples as mentioned above. Furthermore, a failure of the membrane WGC would have important ramifications. Swampland conjectures are intricately connected to one another, and a failure of membrane WGC is sure to have consequences for other cases. For instance, the membrane WGC is related to much better established, and even proven, versions of WGC via T-duality\footnote{It would be interesting to relate a WGC-violating membrane to other cases where violations of WGC have been suggested, such as particles in three-dimensional flat space \cite{Heidenreich:2024dmr,Etheredge:2024amg}.}. Perhaps more importantly, \cite{Ooguri:2016pdq} used the membrane WGC to argue instability of non-supersymmetric AdS, a conjecture that has since been verified in various examples (see e.g. \cite{Freivogel:2016qwc,Banks:2016xpo,Danielsson:2016mtx,Ooguri:2017njy,Bena:2020xxb,Guarino:2020jwv,Henriksson:2019ifu,Suh:2020rma,Apruzzi:2021nle,Basile:2021vxh,Bomans:2021ara,Dibitetto:2022rzy,Naghdi:2020xbi,Suh:2021icf,Marchesano:2021ycx,Casas:2022mnz,Marchesano:2022rpr}). If the membrane WGC turned out to be false, one would be left to wonder why the evidence deceived us so far. It would be interesting to explore these ramifications, a question we leave for the future. 

At any rate, even if the membrane WGC is violated, our results only serve to make the putative dual of the DGKT solution even stranger. On top of the integer conformal dimensions, the rapidly scaling central charge, and the absence of marginal deformations, we have now shown that on a generic point of the moduli space, where the D4 branes that change the $F_4$ flux are in generic position, the theory is non-supersymmetric.  Although the appearance of supersymmetry in non-supersymmetric systems is known to take place in simple low-dimensional quantum systems \cite{Zamolodchikov:1986db,Friedan:1984rv,Lee:2006if}, or even non-holographic 2+1 QFT's, it is a more surprising behavior in a system with an Einstein holographic dual. If the CFT can be engineered by a stack of branes probing a singularity, this must be a non-supersymmetric singularity which becomes supersymmetric when the branes are on top of it -- a quite uncommon behavior.

The whole reason why DGKT is in tension with WGC is because there is no parity symmetry protecting the moduli space of the branes. Similar considerations are likely to apply to classically BPS branes in other 4d $\mathcal{N}=1$ setups, such as the KKLT AdS vacua of \cite{Kachru:2003aw}. Reference \cite{Lust:2022lfc} pointed out that many flux choices in the KKLT scenario cannot be realized by BPS sets of branes; extrapolating our results from DGKT to KKLT, we would conclude that likely \emph{no} flux choice in KKLT can be realized by exactly BPS branes, unless the vacuum preserves some hidden parity symmetry. It is also worth pointing out that this does not apply to the massless IIA version of the DGKT setup, presented in \cite{Cribiori:2021djm}. As discussed in that reference, these solutions, if they exist, uplift to M-theory backgrounds which are completely geometric and controlled, and are protected by a parity symmetry. The only question that needs to be answered is whether the manifolds introduced in \cite{Cribiori:2021djm} admit a weak $G_2$ metric, a question on which recent progress was made \cite{VanHemelryck:2024bas}. We note, however, that the Calabi-Yau version of the question, studied in \cite{Collins:2022nux}, does not encourage optimism.

Both KKLT and DGKT have in common a somewhat involved construction. A much simpler 4d $\mathcal{N}=1$ setup, also involving Romans' mass, is the solution in \cite{Gaiotto:2009yz}\footnote{We thank O. Aharony for bringing up this example.}. Although it is not scale-separated, the solution is in a much better standing, since reference \cite{Gaiotto:2009yz} proposed a concrete  candidate dual CFT, the internal space contains no orientifolds, and there is an explicit ten-dimensional metric. On quick inspection, we could not find an obvious parity symmetry in this system, and we find the possibility of one unlikely, since the dual field theory contains Chern-Simons terms with non-zero total level. If there is no hidden parity symmetry in this setup (e.g. an accidental, emergent parity symmetry in the deep IR), and no other pathology is discovered, we would expect that domain walls get corrected, leading to the same tension with the membrane WGC in AdS as we found in DGKT. We hope to return to this interesting setup in the near future, which is much more amenable to a concrete analysis from both AdS and CFT points of view.

A point we emphasized is that membrane WGC requires an exact moduli space. It is natural to wonder whether the mere existence of a moduli space is enough to imply the absence of scale separation. Some progress in this direction comes from \cite{Cuomo:2024fuy}, which showed the existence of an infinite tower of charged states akin to KK modes, in theories with both a $U(1)$ symmetry and a moduli space. The tower could be very heavy, so this is not directly usable for studies of scale separation, but further improvements might be key to understand this question.

So, all in all: does DGKT exist, or not? We do not know, but perhaps an appealing possibility (or boring, depending on your point of view) is that DGKT does exist as a solution of ten-dimensional massive IIA string theory, but it is rendered non-supersymmetric due to subtle effects. Supersymmetry is very fragile, and it is difficult to rigorously establish it without additional tools such as a dual description. It could be that e.g. non-perturbative effects generate something like  nonsupersymmetric very high-derivative terms in the effective action, while leaving the two-derivative action untouched. We do not have a clue if something like this is actually happening in DGKT but, if it did, the solution would not be protected against e.g. bubble of nothing decays,  and  a lot of the weird features of the holographic dual to DGKT could be ignored. In this respect it is worth noting that we already have scale-separated non-supersymmetric AdS vacua \cite{Luca:2022inb}, which are non-perturbatively unstable and hence do not have a unitary CFT dual.  The general point we wish to emphasize is the fragility of low-supersymmetric systems or vacua, and their sensitivity to UV effects. In particular, one cannot conclude, from the mere fact that the low-energy EFT seems supersymmetric in some approximation, that this is indeed the case, even for parametrically small values of the string coupling and large volume of the compactification manifold.

Prof. Hubert J. Farnsworth claimed in \cite{Futurama} that ``The pursuit of knowledge is hopeless and eternal''.  Although this may be so, we remain hopeful that the more concrete question of the ultimate fate of the DGKT solution can be ascertained in the near future, despite its intricacies, and hope our efforts serve to encourage further work in this important question.

\vspace{0.5cm}

\textbf{Acknowledgements}: We are indebted to Ofer Aharony, Ralph Blumenhagen, Nikolay Bobev, Bruno de Luca, Simeon Hellerman, Ben Heidenreich, Daniel Junghans, Severin L\"{u}st, Juan Maldacena, Fernando Marchesano, Luca Martucci, Liam McAllister, Jakob Moritz, Eran Palti, Kostas Skenderis, Adar Sharon, Alessandro Tomasiello, Angel Uranga, Cumrun Vafa, Max Wiesner and Yoav Zigdon
for enlightening discussions. We are grateful to Ofer Aharony, Daniel Junghans, Fernando Marchesano, Luca Martucci, Eran Palti and especially Angel Uranga for very helpful comments on an early version of this draft. This research was supported in part by grant NSF PHY-2309135 to the Kavli Institute for Theoretical Physics (KITP). The authors thank the KITP program ``Bootstrapping QG'', Harvard University and the Simons Center for Geometry and Physics for hospitality while this project was initiated. We also thank the Erwin Schrodinger International Institute for Mathematics and Physics for their hospitality during the programme ``The Landscape vs. the Swampland''. MM also thanks CERN for hospitality during the completion of this work, as well as the String Phenomenology 2023 conference where an early version of this work was presented and the kind BA employee who allowed unexpected access to an airport lounge. MM is currently supported by the RyC grant RYC2022-037545-I from the AEI and was supported by an Atraccion del Talento Fellowship 2022-
T1/TIC-23956 from Comunidad de Madrid in the early stages of this project. M.M. and I.V. thank the KITP program ``What is String Theory?" for providing a stimulating enviroment for discussion. The authors thank the Spanish Research Agency (Agencia Estatal de Investigacion)
through the grants IFT Centro de Excelencia Severo
Ochoa CEX2020-001007-S and PID2021-123017NB-I00,
funded by MCIN/AEI/10.13039/501100011033 and by
ERDF A way of making Europe. 
The work of I.V. was supported by the grant RYC2019-028512-I from the MCI (Spain), the ERC
Starting Grant QGuide101042568 - StG 2021, and the Project ATR2023-145703 funded by MCIN /AEI /10.13039/501100011033.

 \appendix
 
\section{SCFT of self-intersecting orientifolds in massless IIA}\label{app:A}
This Appendix contains a brief description of the degrees of freedom localized at the self-intersection of two $O6^-$ planes in (massless) IIA string theory. The point is that this is a perfectly ordinary singularity, harboring an SCFT.

Consider type IIA string theory quotiented by the involutions
\begin{equation} (x_7,x_8,x_9)\,\rightarrow (-1)^{F_L}\Omega (-x_7,-x_8,-x_9),\quad (x_5,x_6,x_7,x_8)\,\rightarrow (-x_5,-x_6,-x_7,-x_8).\end{equation}
These describe a pair of transversely intersecting $O6^-$ planes, extending along the directions
\begin{center}\begin{tabular}{c|cccccccccc}
&0&1&2&3&4&5&6&7&8&9\\\hline
$O6$&--&--&--&--&--&--&--&$\times$&$\times$&$\times$\\\hline
$O6$'&--&--&--&--&--&$\times$&$\times$&--&--&$\times$
\end{tabular}\end{center}
This is the only way the intersection can be supersymmetric; the above is a local description of the intersecting orientifold singularities of \cite{Cribiori:2021djm}(see eq. (3.15) of that reference). We focus on these, rather than the benchmark DGKT geometry described in the main text, since the geometry is simpler.  The geometric and NS fluxes described there are perturbations that break to $\mathcal{N}=1$ and that can be neglected in the large flux limit, where the backreactions are small.

We want to understand the strong coupling description of this intersection. The $O6$ planes have 5 directions in common, and the system preserves 5d $\mathcal{N}=1$ supersymmetry. The M theory uplift is simply the $CY_3$ cone defined by\ the intersection in $\mathbb{C}^5$
\begin{equation} z_1z_2=z_3^2,\quad z_4^2=(z_5^2-1)z_2.\label{e23335}\end{equation}
These are simply the equations corresponding to a $\mathbb{C}^2/\mathbb{Z}_2$ singularity and an Atiyah-Hitchin geometry; indeed, the orbifold-orientifold group is generated by the actions corresponding to O6, and O6', or O6 and a flip of the coordinates $(x_5,x_6,x_7,x_8)$. The factor of $-1$ in the equation is associated to the fact that the Atiyah-Hitchin manifold, which is the uplift of a single orientifold, has two $A_1$ singularities \cite{Seiberg:1996nz}. To see that this singularity is indeed the uplift of the intersecting O6's, notice that \eq{e23335} is solved by identifying
\begin{equation} z_1=(x_5+ix_6)^2,\quad z_2=(x_7+ix_8)^2,\quad z_3=(x_5+ix_6)^2(x_7+ix_8)^2\end{equation}
and
\begin{equation} z_4=(x_9+ix_{10})(x_7+ix_8),\quad z_5=\pm\sqrt{1+x_9+ix_{10}}.\end{equation}

We can eliminate $z_2$ in \eq{e23} to obtain a hypersurface in $\mathbb{C}^4$,
\begin{equation} z_1z_4^2=(z_5^2-1)z_3.\label{e24}\end{equation}
This equation has a non-isolated singularity (the singular locus is complex 1-dimensional), which are more difficult to study. To turn it into something more manageable, we can deform it to
\begin{equation} z_1z_4^2+\alpha z_1=(z_5^2-1)z_3,\label{e25}\end{equation}
(which physically corresponds to smoothing out the Atiyah-Hitching geometry of the orientifold),
in which case this is case IV of table 1 in \cite{Xie:2015rpa}, with $a=c=1$, $b=d=2$. When compactified on a circle, this yields a 4d $\mathcal{N}=2$ SCFT of rank 4. 

From a QFT point of view, blowing up the singularity corresponds to giving a vacuum expectation value to some fields, leading to a generic point in the Coulomb branch. The IR theory is free, corresponding to a M theory geometry which has become smooth, and which has localized 2-cycles where gauge fields coming from $C_3$ live.

\bibliographystyle{JHEP}

\bibliography{DGKT-refs}

\end{document}